\documentclass[aps,prd,nofootinbib,onecolumn,a4paper]{revtex4-2}
\usepackage{bm,latexsym,amsmath,amssymb,amsfonts,mathrsfs,amsfonts}
\usepackage[dvipdfmx]{graphicx}
\usepackage[hang,small,bf]{caption}
\usepackage[subrefformat=parens]{subcaption}
\usepackage{ascmac}
\usepackage{physics}
\usepackage{color,float}
\usepackage{braket}
\usepackage{bbm}

\newcommand{\simgt}{\lower.5ex\hbox{$\; \buildrel > \over \sim \;$}}
\newcommand{\simlt}{\lower.5ex\hbox{$\; \buildrel < \over \sim \;$}}

\begin{document}
\title{Generating quantum entanglement between macroscopic objects 
\\
with continuous measurement and feedback control
\\
}

\author{Daisuke Miki,$^{1}$ Nobuyuki Matsumoto,$^{2}$ Akira Matsumura,$^{1}$ Tomoya Shichijo,$^{1}$ \\ Yuuki Sugiyama,$^{1}$ Kazuhiro Yamamoto,$^{1,3}$ Naoki Yamamoto,$^{4,5}$}

\affiliation{$^1$Department of Physics, Kyushu University, 744 Motooka, Nishi-Ku, Fukuoka 819-0395, Japan}

\affiliation{$^2$Department of Physics, Faculty of Science, Gakushuin University, 1-5-1, Mejiro, Toshima, Tokyo, 171-8588 Japan}

\affiliation{
$^3$Research Center for Advanced Particle Physics, Kyushu University, 744 Motooka, Nishi-ku, Fukuoka 819-0395, Japan}

\affiliation{
$^4$Quantum Computing Center, Keio University, Hiyoshi 3-14-1, Kohoku, Yokohama 223-8522, Japan }

\affiliation{
$^5$Department of Applied Physics and Physico-Informatics, Keio University, Hiyoshi 3-14-1, Kohoku, Yokohama 223- 8522, Japan
}
\email{miki.daisuke@phys.kyushu-u.ac.jp, \\
nobuyuki.matsumoto@gakushuin.ac.jp, \\
matsumura.akira@phys.kyushu-u.ac.jp, \\
shichijo.tomoya.351@s.kyushu-u.ac.jp, \\
sugiyama.yuki@phys.kyushu-u.ac.jp,\\
yamamoto@phys.kyushu-u.ac.jp,\\
yamamoto@appi.keio.ac.jp}


\date{\today}
\begin{abstract}
This paper is aimed at investigating the feasibility of generating quantum conditional entanglement between macroscopic mechanical mirrors in optomechanical systems while under continuous measurement and feedback control.
We consider the squeezing of the states of the mechanical common and the differential motions of the mirrors by the action of measuring the common and the differential output light beams in the Fabry-Perot-Michelson interferometer. 
We carefully derive a covariance matrix for the mechanical mirrors in a steady state, employing the Kalman filtering problem with dissipative cavities.
We demonstrate that Gaussian entanglement between the mechanical mirrors is generated when the states of the mechanical common and differential modes of the mirrors are squeezed with high purity in an asymmetric manner.
Our results also show that quantum entanglement between $7$ mg mirrors is achievable in the short term.

\end{abstract}
\maketitle
\section{Introduction}
Cavity optomechanics deals with the coupled dynamics of the oscillating end mirrors of cavities (mechanical oscillators) and the optical mode therein.
This field has the potential to reveal the boundary between the classical and the quantum world \cite{Aspelmeyer,Yambei,QO,Matsumoto,Matsumoto20,MY}.
The quantum states of mechanical oscillators
can be achieved by quantum control through interaction with optical cavity modes, whereas mechanical oscillators lose quantum coherence owing to thermal fluctuations.
The technique of continuous measurement cooling shows the potential to achieve the quantum states of 
macroscopic mechanical oscillators \cite{Matsumoto,MY,Meng}. 
Ref. \cite{Genes} demonstrated cooling a mechanical oscillator to the ground state through cavity detuning and feedback control.
Moreover, optomechanical systems are helpful in generating entanglements.
Ref. \cite{henning} discussed the role of feedback cooling; the authors showed that the entanglement between two levitated nanospheres due to the Coulomb force could be measured experimentally with the feedback-based setup.
The authors in Refs. \cite{Vitali,Miao10} considered the detectability of entanglement between the optical cavity mode and the mechanical oscillator in the ground state.
Refs. \cite{Korppi,Kotler,Lepinay} showed that the generation of quantum entanglement between nanoscale objects was realized experimentally.
Recently, cavity optomechanics has attracted significant interest as a possible field for investigating the quantum nature of gravity through tabletop experiments \cite{Blaushi,Miao,Matsumura,Krisnanda,Datta,Miki2,Plato}.
Entanglement generation due to gravitational interaction can be considered as evidence of the quantum nature of gravity \cite{Bose,MV},
which has sparked several investigations
\cite{Nguyen,Miki,Matsumura21,Sugiyama,Matsumura22,LG,Feng}.
Moreover, related to gravitational entanglement, the quantum nature of gravity has been discussed in gravitons and quantum field theory \cite{Belenchia,Marshman,Carney22,Danielson,Bose22,Hidaka,Sugiyama2,Biswas}.
However, verifying the quantum nature of gravity requires entanglement between heavier objects \cite{Miao,Datta}.
The realization of macroscopic quantum systems is pivotal for investigating the unexplored areas between the quantum world and gravity.

In this paper, we consider the feasibility of realizing 
Gaussian entanglement between macroscopic oscillators via optomechanical coupling.
It is known that entanglement between two squeezed light beams with different squeezing angles is generated by passing them through the beam splitter
(e.g., Ref.~\cite{Furusawa}).
The authors of Ref. \cite{HMullerEbhardt} analyzed the entanglement in a comparable situation where the power-recycled mirror squeezed the oscillators' common and differential modes asymmetrically.
However, their analysis was limited to high-frequency regions, where the oscillators were regarded as free mass. 
Namely, they only demonstrated entanglement generation between Fourier modes of the
macroscopic oscillator's motions in high-frequency regions. 
Therefore the previous work is not enough to include the analysis around resonant frequencies. 
Quantum control of macroscopic oscillators around resonant frequencies is important for entanglement generation (e.g., Refs.~\cite{Miao,Datta}).
Then, our analysis here is not limited to high-frequency regions.

We revisit the realization of entanglement between macroscopic oscillators with the Kalman filter's formalism in a wide range of parameter spaces.
We employ feedback control, which decreases the effective temperature, and detunes enabling us to trap the mechanical oscillator stably with the optical spring, as discussed in Ref. \cite{MY}. 
To clarify the difference between the previous \cite{HMullerEbhardt} and 
present paper, we note that the detuning was not considered in the previous work \cite{HMullerEbhardt}. 
By using these quantum controls in an optomechanical system with a power-recycled mirror, we clarified 
the relationship between the entanglement and 
squeezing of states.
Our results show that
quantum cooperativity and detuning characterize the entanglement behavior, quantum squeezing, and purity.
The entanglement generation requires quantum squeezing of both the common and differential modes of the oscillators.
Squeezing, however,
does not always result in entanglement generation, as high-purity squeezed states are also required.
We demonstrate that the entanglement occurs for the quantum
cooperativity 
$C_{\pm}/n_{\text{th}}^{\pm}\simgt 3$
with the experimentally achievable parameters in amplitude quadrature measurement ($X$ measurement).

The remainder of this paper is organized as follows: In Section II, we present a brief review of 
optomechanical systems while under continuous measurement and feedback control. 
In Section III, we provide a mathematical formula for the Riccati equation to describe the
covariance matrix using a quantum Kalman filter to minimize the correlation. 
In Section IV, we extended the formulations in the previous sections to those with two optomechanical 
systems, in which we consider the entanglement between them through a beam splitter in a power-recycled interferometer. We determined the feasibility of preparing entanglements between the 
mirror oscillators in the space of the model parameter, depending on the amplitude quadrature measurement ($X$ measurement) and phase quadrature measurement ($Y$ measurement), respectively. Finally, Section V presents our conclusions. 
The derivation of the 
input-output relation in interferometer
is presented in Appendix A.
In Appendix B, we describe the details of logarithmic negativity for estimating 
the entanglement developed in this paper.
In Appendix C, we describe the details of computing the squeezing angle.

\section{Formulas}
In this section, we consider a driven optical cavity mode that interacts with an oscillating mirror, which is regarded as a mechanical harmonic oscillator.
The Hamiltonian of our system is as follows:
\begin{align}
    H
    &=\frac{P^{2}}{2m}+\frac{1}{2}m\Omega^{2}Q^{2}+\hbar\omega_{c}a^{\dagger}a+\hbar\frac{\omega_{c}}{\ell}Qa^{\dagger}a+i\hbar E(a^{\dagger}e^{-i\omega_{L}t}-ae^{i\omega_{L}t}),
    \label{hamiltonian}
\end{align}
where $Q$ and $P$ are the canonical position and momentum operators of the oscillator, satisfying the commutation relation $[Q, P]=i\hbar$, while $m$ and $\Omega$ are the mass and resonance frequency 
of the oscillator, respectively; $a$ and $a^{\dagger}$ are the annihilation and creation operators of the optical modes in the cavity, $\ell$ is the cavity length, and $\omega_{c}$ is the cavity frequency.
The last term describes the input laser with frequency $\omega_{L}$ and amplitude $E=\sqrt{P_{\text{in}}\kappa/\hbar\omega_{L}}$, where $P_{\text{in}}$ is the input laser power and $\kappa$ is the optical decay rate.
Here, we introduce non-dimensional variables
\begin{align}
    q
    &=\sqrt{\frac{2m\Omega}{\hbar}}Q,\quad
    p
    =\sqrt{\frac{2}{m\hbar\Omega}}P,
\end{align}
that satisfy the commutation relation $[q, p]=2i$. 

The Langevin equations are given by
\begin{align}
    \dot{q}&=
    \Omega p,\notag\\
    \dot{p}&=
    -\Omega q-2Ga^{\prime\dagger}a'-\Gamma p+\sqrt{2\Gamma}p_{\text{in}},\label{ap:Laeq}\\
    \dot{a}'&=
    i(\omega_{L}-\omega_{c})a'-iGqa'+E-\frac{\kappa}{2}a'+\sqrt{\kappa}a_{\text{in}},\notag
\end{align}
where $a'=e^{i\omega_{L} t}a$ denotes the redefined annihilation operator and $G=(\omega_{c}/\ell)\sqrt{\hbar/2m\Omega}$ is the optomechanical coupling.
$\Gamma$ denotes the mechanical decay rate and $p_{\text{in}}$ is the mechanical noise input with a variance of $\braket{p_{\text{in}}^{2}}=2k_{B}T/\hbar\Omega+1$.
Similarly, $a_{\rm in}$ is the optical noise input 
specified by $\braket{a_{\text{in}}^{2}}=(2N_{\text{th}}+1)/2$ with thermal photon occupation number $N_{\text{th}}$.
Considering the linearization $q\rightarrow \bar{q}+\delta q$, $p\rightarrow \bar{p}+\delta p$, and $a'\rightarrow \bar{a}'+\delta a'$,
we derive the following
equations for the steady state:
\begin{align}
    \dot{\bar{q}}&=
    \Omega \bar{p},\notag\\
    \dot{\bar{p}}&=
    -\Omega \bar{q}-2G|\bar{a}'|^{2}-\Gamma \bar{p},\\
    \dot{\bar{a}}'&=
    i(\omega_{L}-\omega_{c}-G\bar{q})\bar{a}+E-\frac{\kappa}{2}\bar{a}'.\notag
\end{align}
Here, considering 
$\dot{\bar{q}}=\dot{\bar{p}}=\dot{\bar{a}}'=0$,
we have
\begin{align}
    \bar{q}&=
    -2\frac{G}{\Omega}|\bar{a}'|^{2},\notag\\
    \bar{p}&=0,\\
    \bar{a}'&=
    \frac{2E}{\kappa-2i\Delta},\notag
\end{align}
where we define the detuning $\Delta=\omega_{L}-\omega_{c}+2(G|\bar{a}'|)^{2}/\Omega$.
The perturbation equations are as follows:
\begin{align}
    &\dot{\delta q}=
    \Omega\delta p,\\
    \label{p}
    &\dot{\delta p}=
    -\Omega\delta q-2g(e^{-i\phi}\delta a'+e^{i\phi}\delta a^{\prime\dagger})-\Gamma\delta p+\sqrt{2\Gamma}p_{\text{in}}-\int_{-\infty}^{t}ds g_{FB}(t-s)X(s),\\
    &\dot{\delta a}'=
    i\Delta\delta a'-ige^{i\phi}\delta q-\frac{\kappa}{2}\delta
    a'+\sqrt{\kappa}a_{\text{in}},
    \label{a}
\end{align}
where $\bar{a}'=e^{i\phi}|\bar{a}'|$ and $g=(|\bar{a}'|\omega_{c}/\ell)\sqrt{\hbar/2m\Omega}$ denotes the redefined optomechanical coupling.
We add that the last term in Eq. \eqref{p} described the feedback effects \cite{Yambei,MY} and
we henceforth simply represent $(\delta q,\delta p,\delta a')$ as $(q,p,a')$.
By introducing the amplitude quadrature $x=e^{-i\phi}a'+e^{i\phi}a^{\prime\dagger}$ and the 
phase quadrature $y=(e^{-i\phi}a'-e^{i\phi}a^{\prime\dagger})/i$, Eq. (\ref{a}) yields
\begin{align}
    \label{xev}
    \dot{x}
    &=-\frac{\kappa}{2}x-\Delta y+\sqrt{\kappa}x_{\text{in}},\\
    \dot{y}
    &=-\frac{\kappa}{2}y+\Delta x+\sqrt{\kappa}y_{\text{in}}-2gq,
    \label{yev}
\end{align}
where $x_{\text{in}}$ and $y_{\text{in}}$ are the corresponding input noises
similarly defined as $a_{\rm in}$, whose variance is specified by:  
$\braket{x_{\text{in}}^{2}}=\braket{y_{\text{in}}^{2}}=2N_{\text{th}}+1$.

Here, we consider the adiabatic limit $\kappa\gg\Omega$, which allows the continuous measurement of the oscillator position because the cavity photon dissipation is sufficiently larger than the frequency of the oscillator. 
The adiabatic limit
is rephrased as the limit of the dissipation dominant regime where the time derivative term of the optical field $\dot{a}'$ is much smaller than the terms of the right-hand side of Eq.~\eqref{a}. 
Then, the time derivatives of the optical amplitude quadrature $x$ and the phase quadrature $y$ 
are also negligible in Eqs.~\eqref{xev} and \eqref{yev}, which leads to the following equations:
\begin{align}
    x
    &=\frac{8\Delta g}{\kappa^{2}+4\Delta^{2}}q
    +\frac{2\kappa\sqrt{\kappa}}{\kappa^{2}+4\Delta^{2}}x_{\text{in}}
    -\frac{4\Delta\sqrt{\kappa}}{\kappa^{2}+4\Delta^{2}}y_{\text{in}},\\
    y
    &=-\frac{4\kappa g}{\kappa^{2}+4\Delta^{2}}q
    +\frac{4\Delta\sqrt{\kappa}}{\kappa^{2}+4\Delta^{2}}x_{\text{in}}
    +\frac{2\kappa\sqrt{\kappa}}{\kappa^{2}+4\Delta^{2}}y_{\text{in}},
\end{align}
Introducing the rescaled variables 
\begin{align}
    q
    &=q'\sqrt{\frac{\Omega}{\omega_{m}}}, \quad
    p
    =p'\sqrt{\frac{\omega_{m}}{\Omega}}, \quad
    \omega_{m}
    =\sqrt{\Omega^2+\Omega\frac{16\Delta g^2}{\kappa^2+4\Delta^2}},\quad
    g_{m}
    =g\sqrt{\frac{\Omega}{\omega_{m}}},
\end{align}
we rewrite the equation of motion as:
\begin{align}
    \label{qeq}
    \dot{q}'
    &=\omega_{m} p',\\
    \dot{p}'
    &=-\omega_{m} q'-\gamma_{m} p'+\sqrt{2\gamma_{m}}p_{\text{in}}'
    -\frac{4g_{m}\kappa\sqrt{\kappa}}{\kappa^{2}+4\Delta^{2}}x_{\text{in}}
    +\frac{8g_{m}\Delta\sqrt{\kappa}}{\kappa^{2}+4\Delta^{2}}y_{\text{in}},\\
    \label{xeq}
    x
    &=\frac{8\Delta g_{m}}{\kappa^{2}+4\Delta^{2}}q'
    +\frac{2\kappa\sqrt{\kappa}}{\kappa^{2}+4\Delta^{2}}x_{\text{in}}
    -\frac{4\Delta\sqrt{\kappa}}{\kappa^{2}+4\Delta^{2}}y_{\text{in}},\\
    \label{yeq}
    y
    &=-\frac{4\kappa g_{m}}{\kappa^{2}+4\Delta^{2}}q'+\frac{4\Delta\sqrt{\kappa}}{\kappa^{2}+4\Delta^{2}}x_{\text{in}}+\frac{2\kappa\sqrt{\kappa}}{\kappa^{2}+4\Delta^{2}}y_{\text{in}},
\end{align}
where $\gamma_{m}$ is the effective mechanical decay rate under feedback control, and the thermal noise input changes to $\braket{p_{\text{in}}^{\prime2}}=2n_{\text{th}}+1$ with $n_{\text{th}}=k_{B}T\Gamma/\hbar\gamma_{m}\omega_{m}$.

The quadratures of the optical cavity modes $x$ and $y$ contain information regarding the position of the mechanical oscillator $q'$ in Eqs. \eqref{xeq} and \eqref{yeq}.
To estimate the oscillator position, we either consider the measurement of the amplitude quadrature $x$, or the measurement of the phase quadrature $y$.
The amplitude quadrature of the output optical field is obtained by the input-output relation $x_{\text{out}}=x_{\text{in}}-\sqrt{\kappa}x$ \cite{Yambei,Gardiner}.
However, we need to consider the additional noise input due to the imperfect measurement.
Thus, the observation signal of amplitude quadrature $x$ is described by the following equation:
\begin{align}
    X
    &=\sqrt{\eta}x_{\text{out}}+\sqrt{1-\eta}x_{\text{in}}',
\end{align}
where $\eta\in[0,1]$ is the detection efficiency and $x_{\text{in}}'$ is the additional vacuum noise for the imperfect measurement, which satisfies $\langle x_{\rm in}'{}^2\rangle=1$.
Under the limit of the dissipation domination, we have
\begin{align}
    X
    =-\frac{8g_{m}\Delta\sqrt{\eta\kappa}}{\kappa^2+4\Delta^2}q'
    -\sqrt{\eta}\frac{\kappa^{2}-4\Delta^{2}}{\kappa^2+4\Delta^2}x_{\text{in}}
    +\sqrt{\eta}\frac{4\kappa\Delta}{\kappa^2+4\Delta^2}y_{\text{in}}
    +\sqrt{1-\eta}x_{\text{in}}'.
    \label{Xi}
\end{align}
On the other hand, the observation signal of the phase quadrature $y$ is also described by the output equation
\begin{align}
    Y
    &=\sqrt{\eta}y_{\text{out}}+\sqrt{1-\eta}y_{\text{in}}',
    \quad\text{with}\quad
    y_{\text{out}}
    =y_{\text{in}}-\sqrt{\kappa}y,
\end{align}
which reduces to
\begin{align}
    Y
    =\frac{4g_{m}\kappa\sqrt{\eta\kappa}}{\kappa^2+4\Delta^2}q'
    -\sqrt{\eta}\frac{4\kappa\Delta}{\kappa^2+4\Delta^2}x_{\text{in}}
    -\sqrt{\eta}\frac{\kappa^{2}-4\Delta^{2}}{\kappa^2+4\Delta^2}y_{\text{in}}
    +\sqrt{1-\eta}y_{\text{in}}'.
    \label{Yi}
\end{align}

\section{Riccati equation}
Because the observation signals in Eqs. \eqref{Xi} and \eqref{Yi} include noise information, we employed the quantum filter for optimal estimation.
Here, we consider the quantum Kalman filter, which allows us to minimize the mean-squared error
between the canonical operators $\bm{r}=(q',p')^{\text{T}}$ and the estimated values $\tilde{\bm{r}}=(\tilde{q}',\tilde{p})^{\text{T}}$, i.e., each component of the covariance matrix $\bm{V}=\braket{\{\bm{r}-\tilde{\bm{r}},(\bm{r}-\tilde{\bm{r}})^{\text{T}}\}}$ is minimized.
The quantum filter is essential to reduce the thermal fluctuations and increase the squeezing level. 
With the quantum Kalman filter, we can track the behavior of ${\bm r}$ conditioned on the measurement result, and its fluctuation is represented by the conditional covariance matrix following the Riccati equation.
On the other hand, without the quantum filter, we only have the average behavior of $\bm{r}$.
Importantly, the covariance matrix without the filter is always larger than the covariance matrix conditioned on the measurement.
Hence, in the absence of the filter, the squeezing level and, accordingly, the entanglement level must decrease.
This is essential for the entanglement between mechanical mirrors in the next section.

We rewrite the Langevin equation in matrix form as follows:
\begin{align}
    \dot{\bm{r}}
    &=\bm{A}\bm{r}
    +\left(\begin{array}{c}
    0\\
    w
    \end{array}\right),\\
    X
    &=\bm{C}_{X}\bm{r}+v_{X},\label{Xmat}\\
    Y
    &=\bm{C}_{Y}\bm{r}+v_{Y}\label{Ymat},
\end{align}
where,
\begin{align}
    \bm{A}&=
    \left(\begin{array}{cc}
    0&\omega_{m}\\
    -\omega_{m}&-\gamma_{m}
    \end{array}\right),\quad
    w=\sqrt{2\gamma_{m}}p_{in}'
      -\frac{4g_{m}\kappa^{3/2}}{\kappa^{2}+4\Delta^{2}}x_{\text{in}}
      +\frac{8g_{m}\kappa^{1/2}\Delta}{\kappa^{2}+4\Delta^{2}}y_{\text{in}},\\
    \bm{C}_{X}&=
    \left(\begin{array}{cc}
    -\frac{8g_{m}\Delta\sqrt{\eta\kappa}}{\kappa^{2}+4\Delta^{2}}&0
    \end{array}\right),\quad
    v_{X}=
    -\frac{\kappa^{2}-4\Delta^{2}}{\kappa^{2}+4\Delta^{2}}\sqrt{\eta}x_{\text{in}}
    +\frac{4\kappa\Delta}{\kappa^{2}+4\Delta^{2}}\sqrt{\eta}y_{\text{in}}
    +\sqrt{1-\eta}x_{\text{in}}',\\
    \bm{C}_{Y}&=
    \left(\begin{array}{cc}
    \frac{4g_{m}\kappa\sqrt{\eta\kappa}}{\kappa^{2}+4\Delta^{2}}&0
    \end{array}\right),\quad
    v_{Y}=
    -\frac{4\kappa\Delta}{\kappa^{2}+4\Delta^{2}}\sqrt{\eta}x_{\text{in}}
    -\frac{\kappa^{2}-4\Delta^{2}}{\kappa^{2}+4\Delta^{2}}\sqrt{\eta}y_{\text{in}}
    +\sqrt{1-\eta}y_{\text{in}}'.
\end{align}
We use Eq.~\eqref{Xmat} or Eq.~\eqref{Ymat} for the optical amplitude measurement and the phase measurement, respectively.

For the Kalman filter \cite{Yamamoto,Wieczorek},
we obtained the time evolution of the optimized covariance matrix as the following Riccati equation:
\begin{align}
    \label{riccati}
    &\dot{\bm{V}}
    =\bm A\bm V+\bm V\bm A^{\text{T}}+\bm N-(\bm V\bm C^{\text{T}}_{I}+\bm L_{I})M^{-1} (\bm V\bm C^{\text{T}}_{I}+\bm L_{I})^{\text{T}}
    \end{align}
where $I=X$ or $Y$, $M=\braket{v_{X}^{2}}=\braket{v_{Y}^{2}}=(2\eta N_{\text{th}}+1)$,
$\tilde{\bm{r}}$ is the best estimated value to optimized the covariance matrix based on the observation $X$ or $Y$, which follows
    \begin{align}
    &\quad
    \dot{\tilde{\bm{r}}}
    =\bm{A}\tilde{\bm{r}}+(\bm V\bm C^{\text{T}}_{I}+\bm L_{I})M^{-1}(I-C_{I}\tilde{\bm{r}}_{I}),
\end{align}
and each matrix is given by
\begin{align}
    \bm{V}
    &=\left(\begin{array}{cc}
    V_{11}&V_{12}\\
    V_{12}&V_{22}
    \end{array}\right),
    \quad
    \bm N
    =\left(\begin{array}{cc}
    0&0\\
    0&\braket{w^{2}}
    \end{array}\right),\\
    \bm{L}_{X}
    &=\left(\begin{array}{c}
    0\\
    \braket{wv_{X}}
    \end{array}\right),
    \quad
    \bm{L}_{Y}
    =\left(\begin{array}{c}
    0\\
    \braket{wv_{Y}}
    \end{array}\right),
\end{align}
with
\begin{align}
    \braket{w^2}
    &=2\gamma_m(2n_{\text{th}}+1)+\frac{16g_{m}^2\kappa}{\kappa^2+4\Delta^2}(2N_{\text{th}}+1)
    \equiv\bar{n},\\
    \braket{wv_{X}}
    &=\frac{4g_{m}\kappa\sqrt{\kappa\eta}}{\kappa^{2}+4\Delta^{2}}(2N_{\text{th}}+1),\quad
    \braket{wv_{Y}}
    =\frac{8g_{m}\Delta\sqrt{\kappa\eta}}{\kappa^{2}+4\Delta^{2}}(2N_{\text{th}}+1).
\end{align}
In the absence of the quantum filter,
the covariance matrix follows the Lyapunov equation $\dot{\bm{V}}=\bm{A}\bm{V}+\bm{V}\bm{A}^{\text{T}}+\bm{N}$, whose components are always larger than those with the filter.
Hence, we need the quantum filter to suppress the thermal fluctuation.

Considering the steady state $\dot{\bm{V}}=0$, the covariance matrix satisfies the following equation:
\begin{align}
    &2\omega_{m}V_{12}-\lambda_{I}V_{11}^2=0,\nonumber\\
    &(\gamma_{m}+\lambda_{I} V_{11})V_{12}+(V_{11}-V_{22})\omega_{m}+\Lambda_{I}V_{11}=0,\label{veq}\\
    &2\gamma_m V_{22}+2\omega_{m}V_{12}+(\sqrt{\lambda_{I}}V_{12}+\Lambda_{I}/\sqrt{\lambda_{I}})^{2}-\bar{n}=0\nonumber,
\end{align}
where we define
\begin{align}
    \lambda_{X}
    &=\frac{64g_{m}^2\eta\kappa\Delta^{2}}{(2\eta N_{\text{th}}+1)(\kappa^2+4\Delta^2)^2},\quad
    \lambda_{Y}
    =\frac{16g_{m}^2{\eta \kappa^{3}}}{(2\eta N_{\text{th}}+1)(\kappa^2+4\Delta^2)^2},\\
    \Lambda_{X}
    &=-\Lambda_{Y}
    =-\frac{32g_{m}^{2}\eta\kappa^{2}\Delta}{(\kappa^{2}+4\Delta^{2})^{2}}\frac{2N_{\text{th}}+1}{2\eta N_{\text{th}}+1}.
\end{align}
The solution of Eq. \eqref{veq} is derived as follows:
\begin{align}
    V_{11}
    &=\frac{\gamma_{I}-\gamma_{m}}{\lambda_{I}},\notag\\
    V_{12}
    &=\frac{(\gamma_{I}-\gamma_m)^2}{2\lambda_{I}\omega_{m}},
    \label{SolV1}\\
    V_{22}
    &=\frac{(\gamma_{I}-\gamma_m)(2\omega_{m}(\omega_{m}+\Lambda_{I})+\gamma_{I}^{2}-\gamma_{m}\gamma_{I})}{2\lambda_{I}\omega_{m}^2},\notag
\end{align}
where we defined 
\begin{align}
    \gamma_{I}
    &=\sqrt{\gamma_m^2-2\omega_{m}(\omega_{m}+\Lambda_{I})+2\omega_{m}\sqrt{\omega_{m}(\omega_{m}+2\Lambda_{I})+\bar{n}\lambda_{I}}}.
\end{align}
The covariance matrix in Eqs. \eqref{SolV1} is also derived using the Wiener filter for the steady state, and is consistent with Ref. \cite{MY} in relation to the optical amplitude measurement.
The result for the optical phase measurement with $\Delta=0$, $N_{\text{th}}=0$, and with no feedback control
is consistent with that reported in Ref. \cite{Meng}.

\begin{figure}[t]
    \centering
    \includegraphics[width=13.cm]{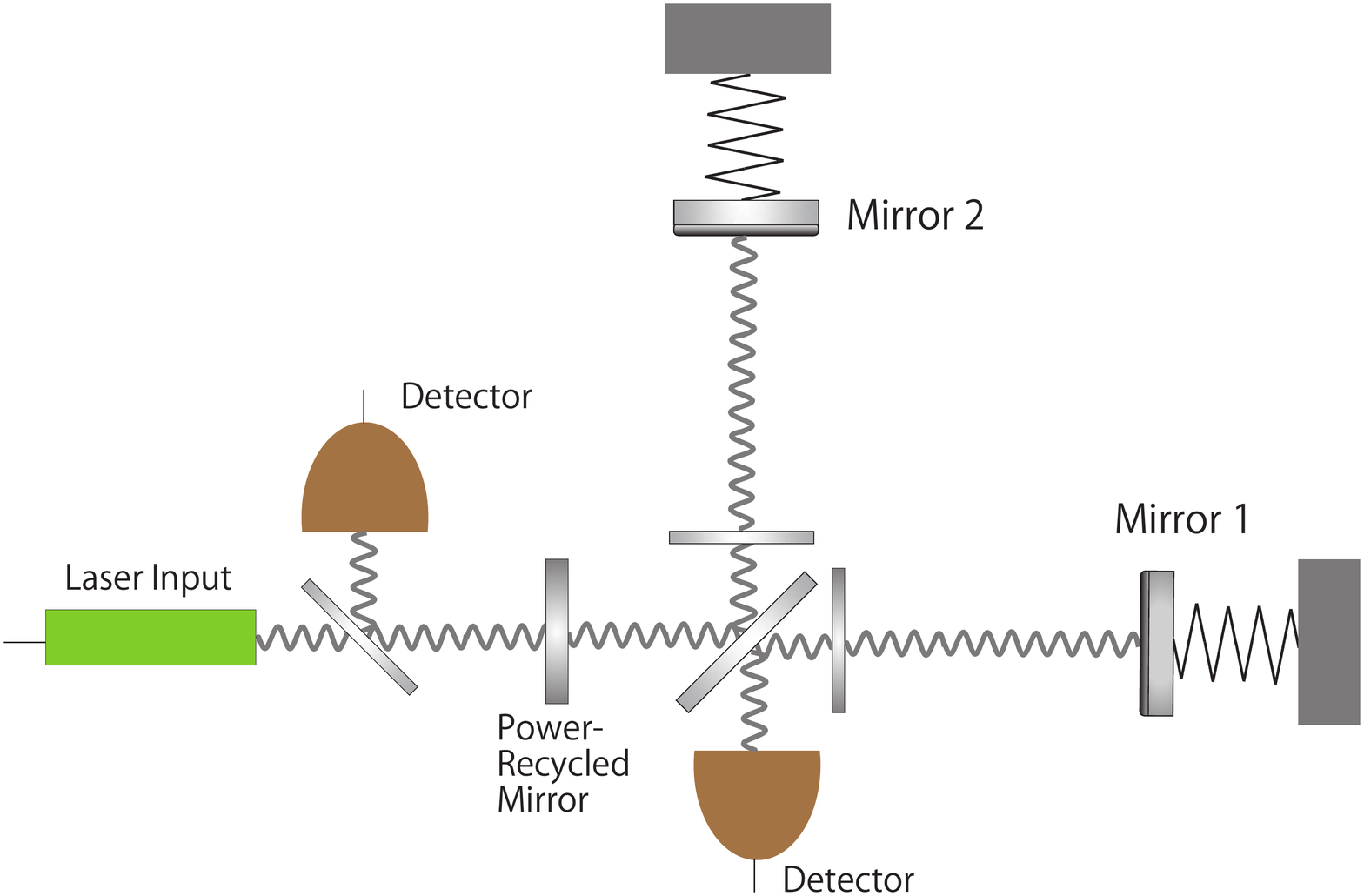}
    \caption{Schematic representation of power-recycled Fabry-P\'{e}rot-Michelson interferometer. 
    }
    \label{fig:PRMI}
\end{figure}

\section{ENTANGLEMENT BETWEEN TWO MIRRORS}
Here, we consider the entanglement between two oscillators coupled with optical modes and passing through the beam splitter in a power-recycled Fabry-P\'{e}rot-Michelson interferometer.
We may consider the case using a signal-recycled mirror (see Refs. \cite{Ebhardt09,Ebhardtphd,Miao12}). However, we only consider the case with the power-recycled mirror for simplicity.
Fig.~\ref{fig:PRMI} shows a schematic plot of this configuration.
In quantum optics, two squeezed beams passing through a half beam splitter become entangled as long as the two squeezed states are not the same (e.g., Ref. \cite{Furusawa}).
Considering entanglement between the mechanical mirrors is analogous to this entanglement generation between the two squeezed beam through a half beam splitter because the optical output quadrature is linearly related to the mechanical mirror position.
Ref. \cite{HMullerEbhardt} shows the entanglement between two oscillators where coupled cavity modes occur by passing the output beams through the beam splitter.
However, the $Y$ measurement is only considered in the high-frequency region, where the oscillator can be considered as a free mass.
In this study, however, our general analysis of the entanglement behavior is not limited to the free-mass region. This is achieved through detuning and feedback effects for $X$ and $Y$ measurements, respectively.

We first consider the Fabry-Perot-Michelson interferometer without the power-recycled mirror.
By introducing the mechanical common and differential modes $q_{\pm}=(q_{1}\pm q_{2})/\sqrt{2}$, $p_{\pm}=(p_{1}\pm p_{2})/\sqrt{2}$, and $a_{\pm}=(a_{1}\pm a_{2})/\sqrt{2}$,
we derive the Langevin equations of the mechanical common and differential modes independently as
\begin{align}
    \dot{q}_{\pm}&=
    \Omega p_{\pm},\notag\\
    \dot{p}_{\pm}&=
    -\Omega q_{\pm}
    -2G(a_{+}^{\prime\dagger}a_{\pm}'\pm a_{-}^{\prime\dagger}a_{\mp}')/\sqrt{2}
    -\Gamma p_{\pm}
    +\sqrt{2\Gamma}p_{\text{in}}^{\pm},\notag\\
    \dot{a}_{+}'&=
    i(\omega_{L}-\omega_{c})a_{+}'-iG(q_{+}a_{+}'+q_{-}a_{-}')/\sqrt{2}
    -\frac{\kappa}{2}a_{+}'
    +\sqrt{\kappa}a_{\text{in}}^{+}
    +E,\label{ap:acd}\\
    \dot{a}_{-}'&=
    i(\omega_{L}-\omega_{c})a_{-}'-iG(q_{+}a_{-}'+q_{-}a_{+}')/\sqrt{2}
    -\frac{\kappa}{2}a_{-}'
    +\sqrt{\kappa}a_{\text{in}}^{-},\notag
\end{align}
where we assume that the individual mirror 1 and mirror 2 follow the same dynamics.
Next, we consider the interferometer shown in Fig. \ref{fig:PRMI} with the power-recycled mirror.
The asymmetry between the mechanical common and differential modes originates from the
power-recycled mirror in the common mode, which is described by the
difference in optical decay rates for each mode $\kappa_\pm$.
The optical decay rates for the differential mode $\kappa_{-}$ and common mode $\kappa_{+}$ are introduced in their 
Langevin equations and the power-recycled mirror makes the optical decay rate of the common mode smaller than that of 
the differential mode by reflecting only a portion of the common-mode light.
Considering the input-output relation of the optical common mode and differential mode,
we introduce the parameter $\zeta\ge1$ to describe the asymmetry using
\begin{equation}
    \label{kcd}
    \kappa_{+}=
    \frac{1}{\zeta}\kappa_{-}.
\end{equation}
We show the details in Appendix A.
Therefore, in our framework, $\zeta$ describes the asymmetry between optical common and differential modes, which 
causes entanglement owing to the half-beam splitter.
Then, the Langevin equations \eqref{ap:acd} are rewritten as
\begin{align}
    \dot{q}_{\pm}&=
    \Omega p_{\pm},\notag\\
    \dot{p}_{\pm}&=
    -\Omega q_{\pm}
    -2G(a_{+}^{\dagger}a_{\pm}'\pm a_{-}^{\prime\dagger}a_{\mp}')/\sqrt{2}
    -\Gamma p_{\pm}
    +\sqrt{2\Gamma}p_{\text{in}}^{\pm},\notag\\
    \dot{a}_{+}'&=
    i(\omega_{L}-\omega_{c})a_{+}'-iG(q_{+}a_{+}'+q_{-}a_{-}')/\sqrt{2}
    -\frac{\kappa_{+}}{2}a_{+}'
    +\sqrt{\kappa_{+}}a_{\text{in}}^{+}
    +E_{+},\label{ap:acd2}\\
    \dot{a}_{-}'&=
    i(\omega_{L}-\omega_{c})a_{-}'-iG(q_{+}a_{-}'+q_{-}a_{+}')/\sqrt{2}
    -\frac{\kappa_{-}}{2}a_{-}'
    +\sqrt{\kappa_{-}}a_{\text{in}}^{-},\notag
\end{align}
where $E_{+}=\sqrt{P_{\text{in}}\kappa_{+}/\hbar\omega_{L}}$ denotes the input-laser amplitude in the common side.
Considering the linearization of the quadratures, we derive the equations for the steady state as
\begin{align}
    \dot{\bar{q}}_{\pm}&=
    \Omega \bar{p}_{\pm}=0,\notag\\
    \dot{\bar{p}}_{\pm}&=
    -\Omega\bar{q}_{\pm}
    -\sqrt{2}G(\bar{a}_{+}^{\prime*}\bar{a}_{\pm}'+\bar{a}_{-}^{\prime*}\bar{a}_{\mp}')
    -\Gamma \bar{p}_{\pm}=0,\notag\\
    \dot{\bar{a}}_{+}'&=
    i(\omega_{L}-\omega_{c})\bar{a}_{+}'
    -iG(\bar{q}_{+}\bar{a}_{+}'+\bar{q}_{-}\bar{a}_{-}')/\sqrt{2}
    -\frac{\kappa_{+}}{2}\bar{a}_{+}'
    +E_{+}=0,\label{ap:bg2}\\
    \dot{\bar{a}}_{-}'&=
    i(\omega_{L}-\omega_{c})\bar{a}_{-}'
    -iG(\bar{q}_{+}\bar{a}_{-}'+\bar{q}_{-}\bar{a}_{+}')/\sqrt{2}
    -\frac{\kappa_{-}}{2}\bar{a}_{-}'=0.\notag
\end{align}
Assuming that the individual quadratures are equal to $\bar{q}_{1}=\bar{q}_{2}$, $\bar{p}_{1}=\bar{p}_{2}$, and $\bar{a}_{1}'=\bar{a}_{2}'$, the quadratures of the differential mode are zero $\bar{q}_{-}=\bar{p}_{-}=\bar{a}_{-}'=0$.
Hence, equation \eqref{ap:bg2} can be rewritten as
\begin{align}
    \bar{q}_{+}&=\sqrt{2}\bar{q}_{1}=
    -\frac{\sqrt{2}G|\bar{a}_{+}'|^{2}}{\Omega},\quad
    \bar{q}_{-}=0,\notag\\
    \bar{p}_{\pm}&=0,\\
    \bar{a}_{+}'&=\sqrt{2}\bar{a}_{1}'=
    \frac{2E_{+}}{\kappa_{+}-2i\Delta},\quad
    \bar{a}_{-}'=0,\notag
\end{align}
where we define the detuning as:
\begin{align}
    \Delta&=
    \omega_{L}-\omega_{c}-\frac{G}{\sqrt{2}}\bar{q}_{+}\notag\\
    &=\omega_{L}-\omega_{c}+2\frac{G^{2}}{\Omega}|\bar{a}_{1}'|^{2}.
\end{align}
Then, the perturbation equations are:
\begin{align}
    \dot{q}_{\pm}&=
    \Omega p_{\pm},\notag\\
    \dot{p}_{\pm}&=
    -\Omega q_{\pm}
    -2g(e^{-i\phi}a_{\pm}+e^{i\phi}a_{\pm}^{\prime\dagger})
    -\Gamma p_{\pm}+\sqrt{2\Gamma}p_{\text{in}}^{\pm},\\
    \dot{a}_{\pm}'&=
    i\Delta a_{\pm}'
    -ige^{i\phi}q_{\pm}
    -\frac{\kappa_{\pm}}{2}a_{\pm}'
    +\sqrt{\kappa_{\pm}}a_{\text{in}}^{\pm},\notag
\end{align}
where we denote the quadratures $(\delta q_{\pm},\delta p_{\pm},\delta a'_{\pm})$ as $(q_{\pm},p_{\pm},a'_{\pm})$.
The optomechanical coupling is
\begin{align}
    g&=
    |\bar{a}_{+}'|G/\sqrt{2}\notag\\
    &=|\bar{a}_{1}'|G,
\end{align}
and $\bar{a}_{+}'=|\bar{a}_{+}'|e^{i\phi}$.

Considering the adiabatic limit $\kappa_{\pm}\gg\Omega$, we derive the equation of motions in the same form of Sec. II as
\begin{align}
    \label{qeq}
    \dot{q}'_{\pm}
    &=\omega_{m} p_{\pm}',\\
    \dot{p}_{\pm}'
    &=-\omega_{m}^{\pm} q_{\pm}'-\gamma_{m} p_{\pm}'+\sqrt{2\gamma_{m}}p_{\text{in}}^{\prime\pm}
    -\frac{4g_{m}^{\pm}\kappa_{\pm}\sqrt{\kappa_{\pm}}}{\kappa_{\pm}^{2}+4\Delta^{2}}x_{\text{in}}^{\pm}
    +\frac{8g_{m}^{\pm}\Delta\sqrt{\kappa_{\pm}}}{\kappa_{\pm}^{2}+4\Delta^{2}}y_{\text{in}}^{\pm},\\
    \label{xeq}
    x_{\pm}
    &=\frac{8\Delta g_{m}^{\pm}}{\kappa_{\pm}^{2}+4\Delta^{2}}q_{\pm}'
    +\frac{2\kappa_{\pm}\sqrt{\kappa_{\pm}}}{\kappa_{\pm}^{2}+4\Delta^{2}}x_{\text{in}}^{\pm}
    -\frac{4\Delta\sqrt{\kappa_{\pm}}}{\kappa_{\pm}^{2}+4\Delta^{2}}y_{\text{in}}^{\pm},\\
    \label{yeq}
    y_{\pm}
    &=-\frac{4\kappa g_{m}^{\pm}}{\kappa_{\pm}^{2}+4\Delta^{2}}q_{\pm}'+\frac{4\Delta\sqrt{\kappa_{\pm}}}{\kappa_{\pm}^{2}+4\Delta^{2}}x_{\text{in}}^{\pm}+\frac{2\kappa_{\pm}\sqrt{\kappa_{\pm}}}{\kappa_{\pm}^{2}+4\Delta^{2}}y_{\text{in}}^{\pm},
\end{align}
where
\begin{align}
    q_{\pm}
    &=q_{\pm}'\sqrt{\frac{\Omega}{\omega_{m}^{\pm}}}, \quad
    p_{\pm}
    =p_{\pm}'\sqrt{\frac{\omega_{m}^{\pm}}{\Omega}}, \quad
    \omega_{m}^{\pm}
    =\sqrt{\Omega^2+\Omega\frac{16\Delta g^2}{\kappa_{\pm}^2+4\Delta^2}},\quad
    g_{m}^{\pm}
    =g\sqrt{\frac{\Omega}{\omega_{m}^{\pm}}}.
\end{align}
$x_{\pm}$ and $y_{\pm}$ are the optical amplitude and phase quadratures.
$\gamma_{m}$ is the effective mechanical decay rate under feedback control, the optical noise input is $\braket{(x_{\text{in}}^{\pm})^{2}}=\braket{(y_{\text{in}}^{\pm})^{2}}=2N_{\text{th}}+1$, and the thermal noise input is $\braket{(p_{\text{in}}^{\prime\pm})^{2}}=2n_{\text{th}}^{\pm}+1$ with $n_{\text{th}}^{\pm}=k_{B}T\Gamma/\hbar\gamma_{m}\omega_{m}^{\pm}$.

We consider the measurement of the optical amplitude quadratures of the common mode and the differential mode.
Due to the imperfect detection, the output quadratures are
\begin{align}
    \label{Xcd}
    X_{\pm}&=
    \sqrt{\eta}x_{\text{out}}^{\pm}+\sqrt{1-\eta}x_{\text{in}}^{\pm\prime},
\end{align}
where $\eta\in[0,1]$ is the detection efficiency and the additional vacuum noise is $\braket{(x_{\text{in}}^{\pm\prime})^{2}}=1$.
Under the adiabatic limit, we have
\begin{align}
\label{Xpm}
    X_{\pm}
    =-\frac{8g_{m}^{\pm}\Delta\sqrt{\eta\kappa_{\pm}}}{\kappa_{\pm}^2+4\Delta^2}q_{\pm}'
    -\sqrt{\eta}\frac{\kappa_{\pm}^{2}-4\Delta^{2}}{\kappa_{\pm}^2+4\Delta^2}x_{\text{in}}^{\pm}
    +\sqrt{\eta}\frac{4\kappa_{\pm}\Delta}{\kappa_{\pm}^2+4\Delta^2}y_{\text{in}}^{\pm}
    +\sqrt{1-\eta}x_{\text{in}}^{\pm\prime}.
\end{align}
For the optical phase quadratures measurement, we similarly obtain
\begin{align}
\label{Ypm}
    Y_{\pm}
    =\frac{4g_{m}^{\pm}\kappa_{\pm}\sqrt{\eta\kappa_{\pm}}}{\kappa_{\pm}^2+4\Delta^2}q_{\pm}'
    -\sqrt{\eta}\frac{4\kappa_{\pm}\Delta}{\kappa_{\pm}^2+4\Delta^2}x_{\text{in}}^{\pm}
    -\sqrt{\eta}\frac{\kappa_{\pm}^{2}-4\Delta^{2}}{\kappa_{\pm}^2+4\Delta^2}y_{\text{in}}^{\pm}
    +\sqrt{1-\eta}y_{\text{in}}^{\pm\prime}.
\end{align}

Then, we consider the Kalman filter to optimize the covariance matrix of the mechanical common mode and differential mode.
Here, there is no correlation between the mechanical common mode and differential mode since the common mode is commutative with the differential mode.
Using the Riccati equation \eqref{riccati} for the steady state,
we obtain the components of the mechanical covariance matrices
$\bm{V}_{\pm}$ as follows:
\begin{align}
    &V_{11}^{\pm}
    =\frac{\gamma_{I}^{\pm}-\gamma_{m}}{\lambda_{I}^{\pm}},~~~~~~\notag\\
    &V_{12}^{\pm}
    =\frac{(\gamma_{I}^{\pm}-\gamma_m)^2}{2\lambda_{I}^{\pm}\omega_{m}^{\pm}},
    \label{cvcd}\\
    &V_{22}^{\pm}
    =\frac{(\gamma_{I}^{\pm}-\gamma_{m})(2\omega_{m}^{\pm}(\omega_{m}^{\pm}+\Lambda_{I}^{\pm})+(\gamma_{I}^{\pm})^{2}-\gamma_{m}\gamma_{I}^{\pm})}{2\lambda_{I}^{\pm}(\omega_{m}^{\pm})^2},\notag
\end{align}
where
\begin{align}
    \lambda_{X}^{\pm}
    &=\frac{64(g_{m}^{\pm})^2\kappa_{\pm}\Delta^{2}\eta}{(2\eta N_{\text{th}}+1)(\kappa_{\pm}^2+4\Delta^2)^2},\quad
    \lambda_{Y}^{\pm}
    =\frac{16(g_{m}^{\pm})^2\kappa_{\pm}^{3}\eta}{(2\eta N_{\text{th}}+1)(\kappa_{\pm}^2+4\Delta^2)^2},\\
    \Lambda_{X}^{\pm}
    &=-\Lambda_{Y}^{\pm}
    =-\frac{32(g_{m}^{\pm})^{2}\kappa_{\pm}^{2}\Delta\eta}{(\kappa_{\pm}^{2}+4\Delta^{2})^{2}}\frac{2N_{\text{th}}+1}{2\eta N_{\text{th}}+1},\\
    \bar{n}_{\pm}
    &=2\gamma_{m}(2n_{\text{th}}^{\pm}+1)+\frac{16(g_{m}^{\pm})^2\kappa_{\pm}}{\kappa_{\pm}^2+4\Delta^2}(2N_{\text{th}}+1),\quad
    n_{\text{th}}^{\pm}
    =\frac{k_{B}T\Gamma}{\hbar\gamma_{m}\omega_{m}^{\pm}},\\
    \gamma_{I}^{\pm}
    &=\sqrt{\gamma_m^2-2\omega_{m}^{\pm}(\omega_{m}^{\pm}+\Lambda_{I}^{\pm})+2\omega_{m}^{\pm}\sqrt{\omega_{m}^{\pm}(\omega_{m}^{\pm}+2\Lambda_{I}^{\pm})+\bar{n}_{\pm}\lambda_{I}^{\pm}}}.
\end{align}

Then, we obtain the solution for each oscillator's canonical operator with a transformation operation using the half-beam splitter:
\begin{align}
    \left(\begin{array}{c}
    \mathcal{Q}_{1}\\
    \mathcal{P}_{1}\\
    \mathcal{Q}_{2}\\
    \mathcal{P}_{2}
    \end{array}\right)
    &=\bm{S}
    \left(\begin{array}{c}
    \mathcal{Q}_{+}\\
    \mathcal{P}_{+}\\
    \mathcal{Q}_{-}\\
    \mathcal{P}_{-}
    \end{array}\right),
    \quad
    \bm{S}
    =\frac{1}{\sqrt{2}}
    \left(\begin{array}{cccc}
    1&0&1&0\\
    0&1&0&1\\
    1&0&-1&0\\
    0&1&0&-1
    \end{array}\right),
\end{align}
where $\mathcal{Q}_{j}$ and $\mathcal{P}_{j}$ with $j=1,2$ denote the dimensional position operator and momentum operator for each mirror, and $\mathcal{Q}_{+}$ and $\mathcal{P}_{+}$ 
($\mathcal{Q}_{-}$ and $\mathcal{P}_{-}$) denote the dimensional position operator and momentum operator for the common mode (the differential mode), satisfying $[\mathcal{Q}_{\pm},\mathcal{P}_{\pm}]=i\hbar$.
The covariance matrix with the basis of the individual mirror $(\mathcal{Q}_1,\mathcal{P}_1,\mathcal{Q}_2,\mathcal{P}_2)$ is given by
\begin{align}
    \bm{\mathcal{V}}
    &=\bm{S}
    \left(
    \begin{array}{cc}
    \bm{\mathcal{V}}_{+}&0\\
    0&\bm{\mathcal{V}}_{-}
    \end{array}
    \right)
    \bm{S}
    \equiv
    \left(
    \begin{array}{cc}
    \bm{\mathcal{V}}_{1}&\bm{\mathcal{V}}_{12}\\
    \bm{\mathcal{V}}_{12}&\bm{\mathcal{V}}_{2}
    \end{array}
    \right),
    \label{vprime}
\end{align}
where $\bm{\mathcal{V}_{\pm}}$ is the covariance matrix with the basis $(\mathcal{Q}_{\pm},\mathcal{P}_{\pm})$ defined as
\begin{align}
    \bm{\mathcal{V}}_\pm
    &=
    \left(\begin{array}{cc}
    \frac{\hbar}{2m\omega_{m}^{\pm}}V_{11}^\pm&\frac{\hbar}{2}V_{12}^\pm\\
    \frac{\hbar}{2}V_{12}^\pm&\frac{2}{m\hbar\omega_{m}^{\pm}}V_{22}^{\pm}
    \end{array}\right),
    \label{primeVpm}
    \end{align}
and $\bm{\mathcal{V}_{1}}$, $\bm{\mathcal{V}}_{12}$, and $\bm{\mathcal{V}_{2}}$ are 
$2\times2$ component matrices.
To analyze the entanglement behavior
between the individual mirror 1 and mirror 2 in Fig. \ref{fig:PRMI},
we introduce logarithmic negativity \cite{Krisnanda,Ebhardt09} with the basis $(\mathcal{Q}_{1},\mathcal{P}_{1},\mathcal{Q}_{2},\mathcal{P}_{2})$ as
\begin{align}
    \label{EN}
    E_{N}
    &=\text{max}\left\{
    0,-\log_{2}\left(\frac{2}{\hbar}\sqrt{\frac{\Sigma-\sqrt{\Sigma^{2}-4{\rm det}\bm{\mathcal{V}}}}{2}}\right)
\right\},
\end{align}
where $\Sigma=\text{det}\bm{\mathcal{V}}_{1}+\text{det}\bm{\mathcal{V}}_{2}-2\text{det}\bm{\mathcal{V}}_{12}$.
For a two-mode Gaussian state, the system is only entangled if the logarithmic negativity is positive $E_{N}>0$.
The critical value $\epsilon_{\text{cr}}$ is defined as the second part of the brace in Eq. \eqref{EN},
\begin{align}
    \epsilon_{\text{cr}}
    &=-\log_{2}\left(\frac{2}{\hbar}\sqrt{\frac{\Sigma-\sqrt{\Sigma^{2}-4{\rm det}\bm{\mathcal{V}}}}{2}}\right)
\end{align}
and $\epsilon_{\text{cr}}>0$ shows that the state is entangled.
Using the steady-state covariance matrix for the mechanical common and differential modes, we have
\begin{align}
    \label{Sigma}
    &\Sigma
    =\frac{\hbar^{2}(\gamma_{+}-\gamma_{m})(\gamma_{-}-\gamma_{m})}{8\lambda_{I}^{+}\lambda_{I}^{-}}
    \left(
    \frac{\gamma_{+}^{2}+\gamma_{-}^{2}-\gamma_{+}\gamma_{-}-\gamma_{m}^{2}}{\omega_{m}^{+}\omega_{m}^{-}}
    +2\frac{\Lambda_{I}^{+}+\omega_{m}^{+}}{\omega_{m}^{-}}
    +2\frac{\Lambda_{I}^{-}+\omega_{m}^{-}}{\omega_{m}^{+}}
    \right),\\
    \label{detV}
    &\text{det}\bm{\mathcal{V}}
    =\left(\frac{\hbar^{2}(\gamma_{+}-\gamma_{m})(\gamma_{-}-\gamma_{m})}{16\lambda_{I}^{+}\lambda_{I}^{-}}\right)^{2}
    \left(\frac{\gamma_{+}^{2}-\gamma_{m}^{2}}{(\omega_{m}^{+})^{2}}+4\frac{\Lambda_{I}^{+}}{\omega_{m}^{+}}+4\right)
    \left(\frac{\gamma_{-}^{2}-\gamma_{m}^{2}}{(\omega_{m}^{-})^{2}}+4\frac{\Lambda_{I}^{+}}{\omega_{m}^{-}}+4\right).
\end{align}

We introduce the quality factor $Q_{\pm}$ and cooperativity $C_{\pm}$ as:
\begin{align}
    \label{QC}
    Q_{\pm}
    &=\frac{\omega_{m}^{\pm}}{\gamma_{m}},\quad
    C_{\pm}
    =\frac{4(g^{\pm}_{m})^{2}}{\gamma_{m}\kappa_{\pm}}.
\end{align}
We can derive logarithmic negativity as a function of $Q_{\pm}$, $C_{\pm}$, $n_{\text{th}}^{\pm}$, $N_{\text{th}}$, $\Delta/\kappa_{\pm}$, and $\eta$ (see Appendix \ref{LN}). Additionally, our results are not limited to the 
free-mass region.
By introducing the normalized detuning $\delta_{\pm}=\Delta/\kappa_{\pm}$, the relation \eqref{kcd} leads to:
\begin{align}
    \label{Q+}
    Q_{+}&=
    Q_{-}\left(
    1+\frac{4C_{-}\delta_{-}(\zeta^{2}-1)}{Q_{-}(1+4\delta_{-}^{2})(1+4\zeta^{2}\delta_{-}^{2})}
    \right)^{1/2},\\
    C_{+}&=
    \zeta C_{-}\left(
    1+\frac{4C_{-}\delta_{-}(\zeta^{2}-1)}{Q_{-}(1+4\delta_{-}^{2})(1+4\zeta^{2}\delta_{-}^{2})}
    \right)^{-1/2} .
    \label{C+}
\end{align}
Whereas the resonance frequency is written as $\Omega/\gamma_{m}=\sqrt{Q(Q-4C\delta+4Q\delta^{2})/(1+4\delta^{2})}$, the stability condition $Q_{\pm}(1+4\delta_{\pm}^{2})>4C_{\pm}\delta_{\pm}$ is always satisfied for $\delta_{\pm}>0$ from the definition \eqref{QC}.
As a result, we note that the stability condition is satisfied for any $(C_{\pm},Q_{\pm},\delta_{\pm})$ as long as $\delta_{\pm}>0$.

\begin{table}[t]
  \centering
  \begin{tabular}{lccr}
    \hline
    Symbol  & Name  &  Value  &  Reference\\
    \hline
    $\Omega$ & Mechanical frequency & $2\pi\times2.2~\text{Hz}$ & \cite{Matsumoto20,MY}\\
    $\Gamma(\Omega)$ & Mechanical decay rate & $2\pi\times10^{-6}~\text{Hz}$ & \cite{Matsumoto20}\\
    $T$ & Bath temperature & $300~\text{K}$ & \cite{Matsumoto20,MY}\\
    $\eta$ & Detection efficiency & $0.92$ & \cite{Matsumoto20,MY}\\
    $m$ & Mirror mass & $7.71\times10^{-6}~\text{kg}$ & \cite{Matsumoto20,MY}\\
    $\ell$ & Cavity length & $10^{-1}~\text{m}$ & \cite{Matsumoto20,MY}\\
    $\omega_{L}(\simeq\omega_{c})$ & Laser frequency & $2\pi \times 300\times10^{12}~\text{Hz}$ & \cite{Matsumoto20,MY}\\
    $\kappa_{-}$ & Optical decay rate & $2\pi\times1.64\times10^{6}~\text{Hz}$ & \cite{Matsumoto20,MY}\\
    $F=2\pi c/\ell\kappa_{-}$ & Finesse & $1.8\times10^{3}$ & \cite{Matsumoto20,MY}\\
    $P_{in}$ & Input laser power & $30~\text{mW}$ & \cite{Matsumoto20,MY}\\
    $\gamma_{m}$ & Effective mechanical decay rate under feedback control & $2\pi\times6.9\times10^{-3}~\text{Hz}$ & \\
    $N_{\text{th}}$ & Thermal photon number & $0$ & \\
    $\delta_{-}=\Delta/\kappa_{-}$ & (Normalized) detuning & $0.2$ & \\
    $\zeta$ & Normalized detuning ratio of differential mode to common mode & $3$ & \\
    \hline
    $|\bar{a}|$ & Expectation value of cavity photon quadrature & $1.27\times10^{5}$ & \\
    $g=|\bar{a}|(\omega_{c}/\ell)\sqrt{\hbar/2m\Omega}$ & Optomechanical coupling & $2\pi\times2.68\times10^{5}~\text{Hz}$ & \\
    $Q_{-}=\omega_{m}^{-}/\gamma_{m}$ & Quality factor & $7.5\times10^{4}$ & \\
    $Q_{+}$ is defined by Eq.~(\ref{Q+})& & $1.6\times10^{5}$ & \\
    $C_{-}=4(g_{m}^{-})^{2}/\gamma_{m}\kappa_{-}$ & Cooperativity & $1.1\times10^{5}$ & \\
    $C_{+}$ is defined by Eq.~(\ref{C+}) & & $1.6\times10^{5}$ & \\
    $n_{\text{th}}^{-}$ & Thermal phonon number & $7.5\times10^{3}$ & \\
    $n_{\text{th}}^{+}$~is defined by Eq.~(\ref{nth+}) & & $1.8\times10^{3}$ & \\
    \hline
  \end{tabular}
  \caption{
  Parameters employed in Ref. \cite{Matsumoto20,MY} and expected from the  state-of-art technique.
  The detuning for the optical common mode is $\delta_{+}=\zeta \delta_{-}$, and  $Q_{+}$, $C_{+}$, and $n_{\text{th}}^{+}$ are decided in Eqs. \eqref{Q+}-\eqref{nth+}.
  }
  \label{tab:parameter}
\end{table}

\begin{figure}[t]
    \centering
    \includegraphics[width=8.2cm]{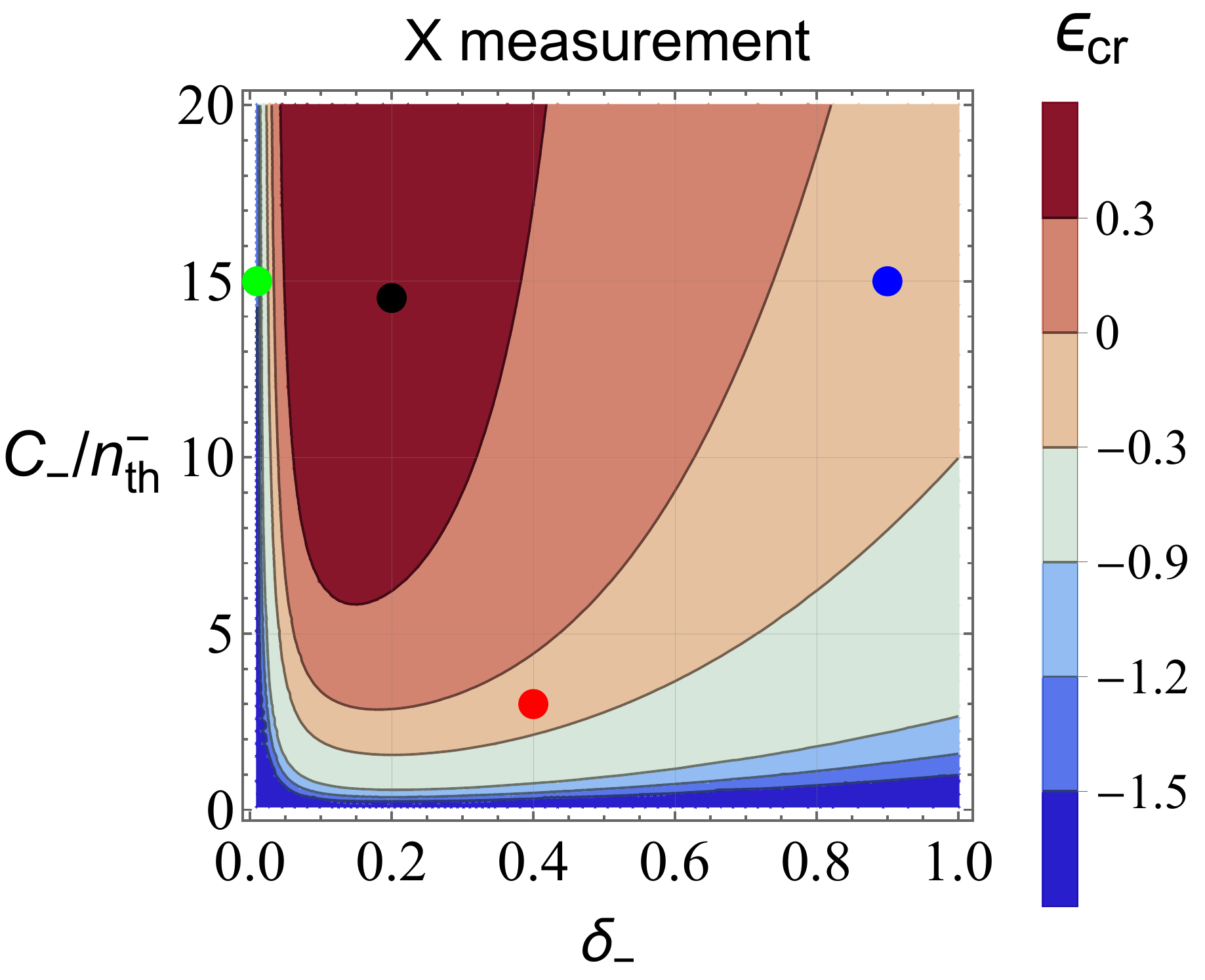}
    \hspace{1.cm}
    \includegraphics[width=7.5cm]{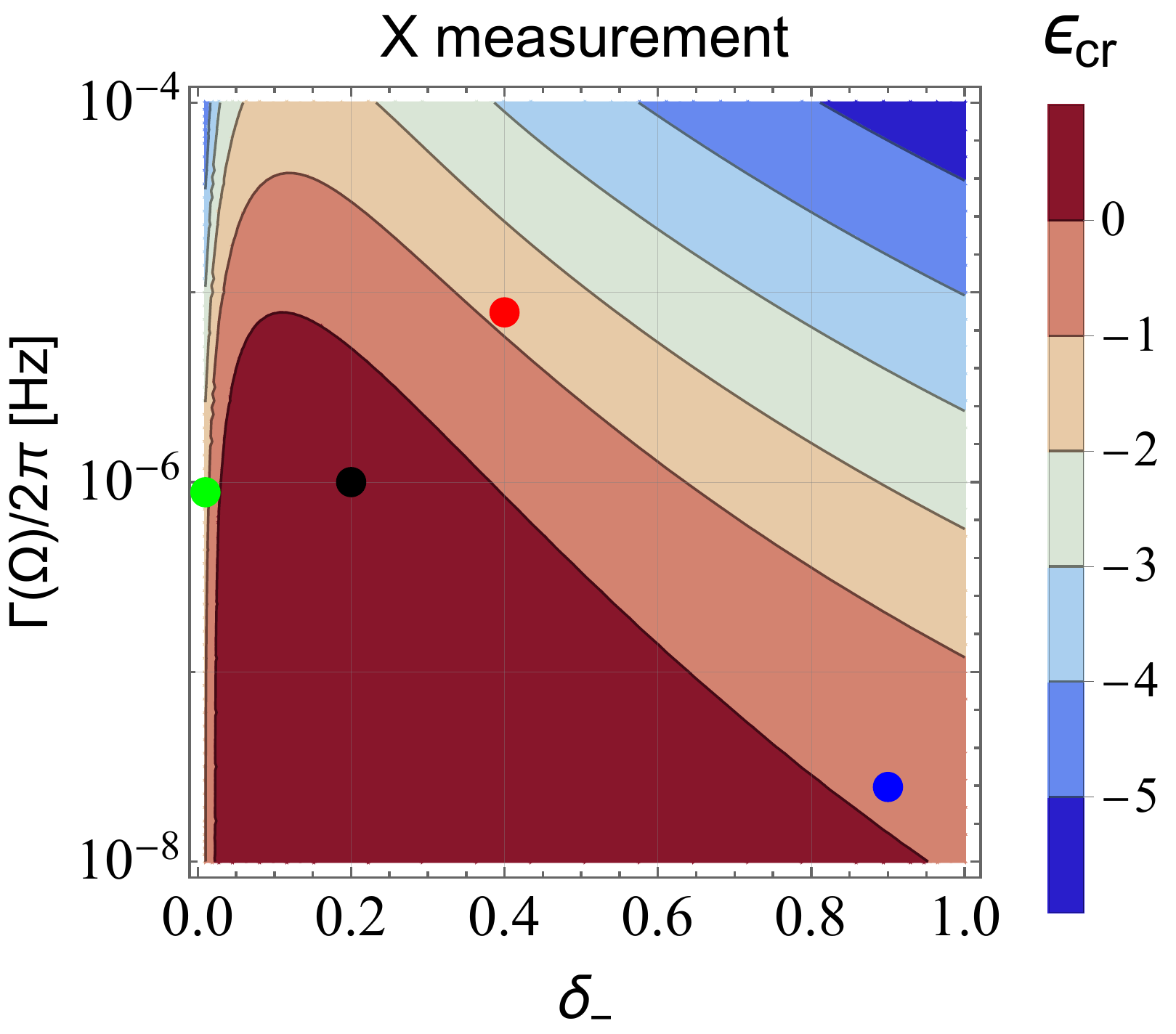}\\
    \vspace{0.5cm}
    \includegraphics[width=6.5cm]{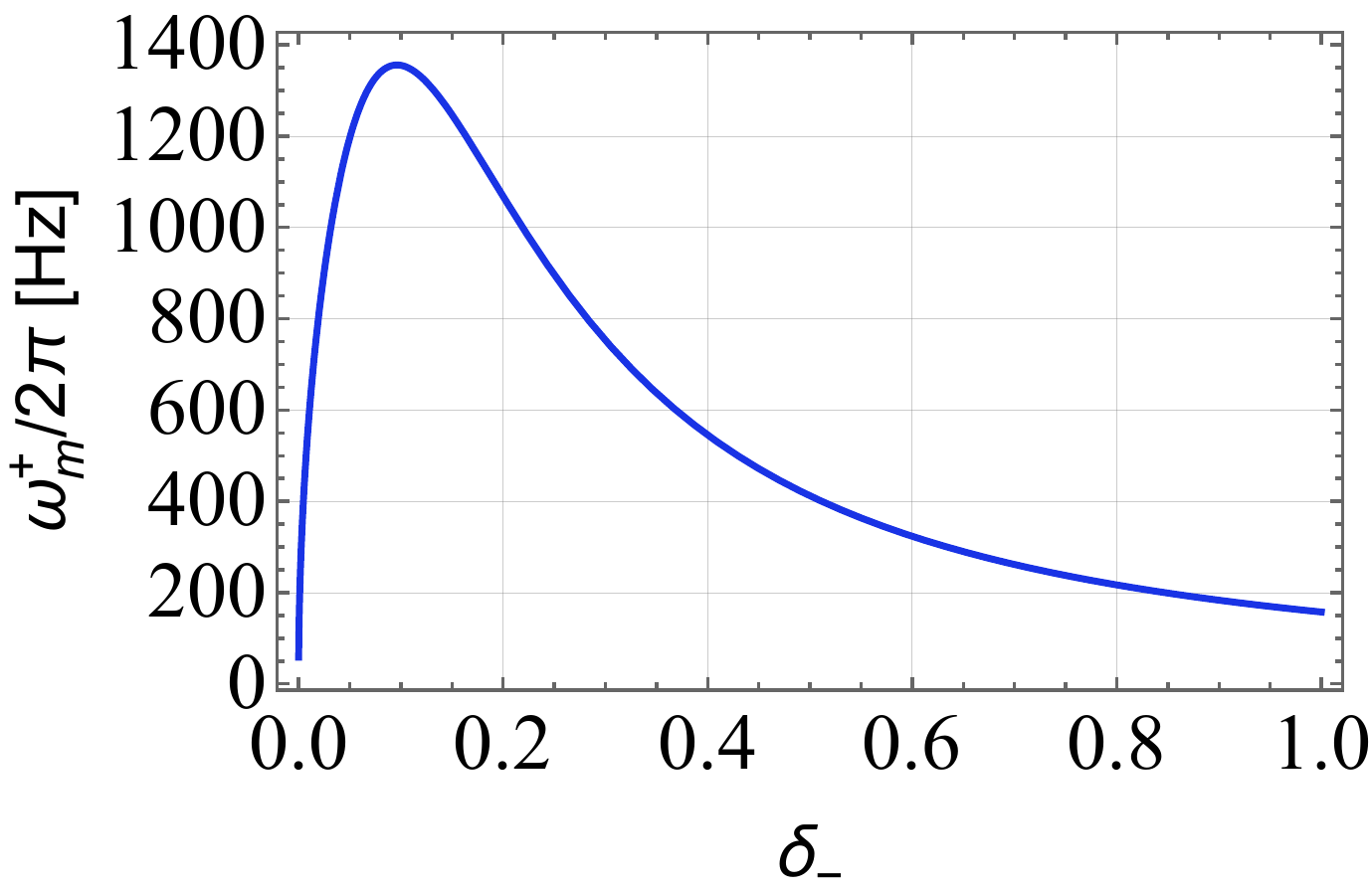}
    \hspace{1.0cm}
    \includegraphics[width=6.5cm]{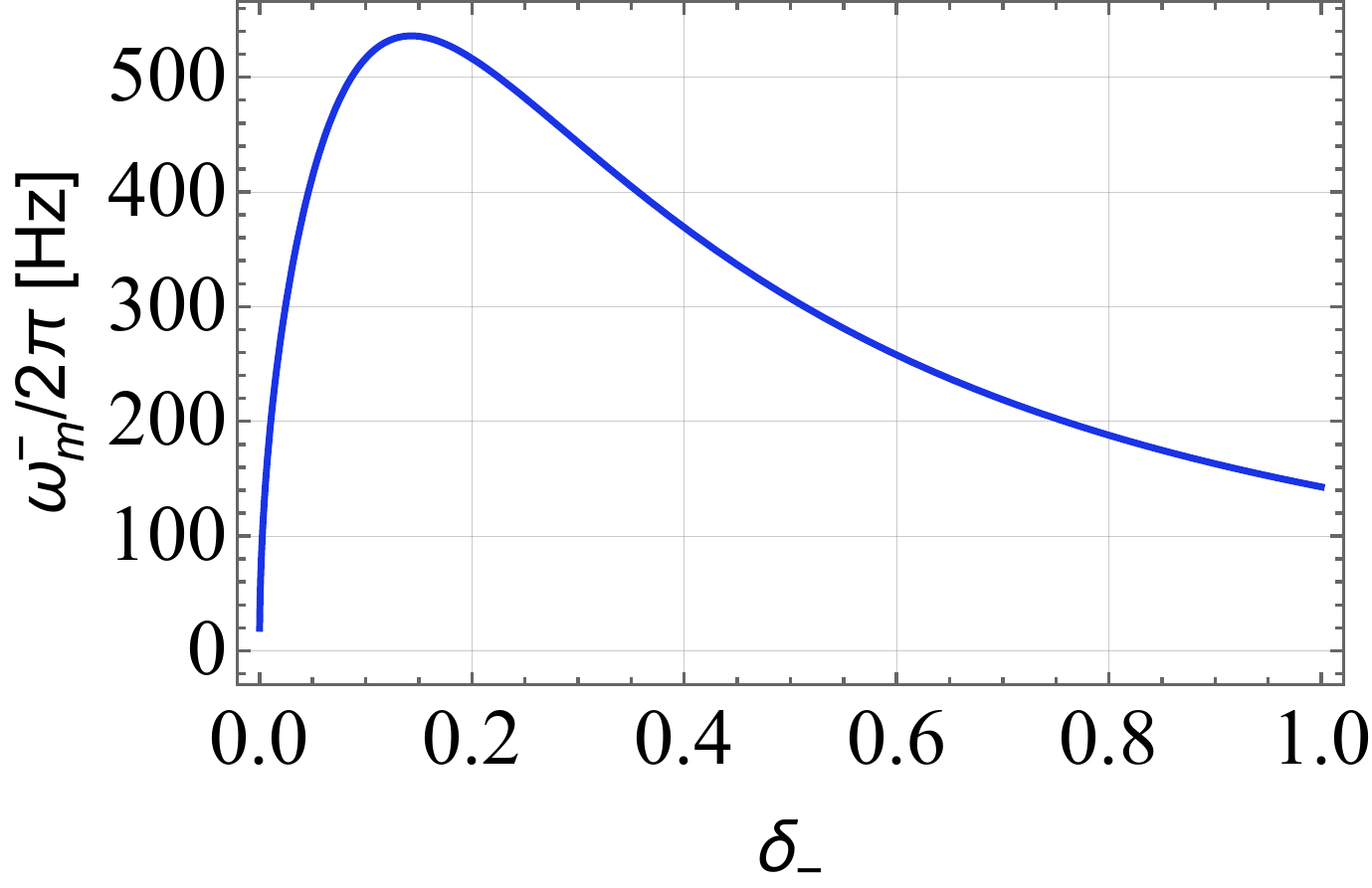}\\
    \caption{Upper panels:
    The critical value $\epsilon_{\rm cr}$ as a function of $C_{-}/n_{\text{th}}^{-}$ and $\delta_-$
    is shown in the upper left panel, while the same is shown as a function of $\Gamma(\Omega)/2\pi$ and $\delta_-$ in the upper right panel.
    We consider the $X$ measurement of both the optical amplitude quadratures of the common and differential modes $X_{\pm}$.
    We assumed the structural damping $\Gamma(\omega_{m}^{\pm})=\Gamma(\Omega)\Omega/\omega_{m}^{\pm}$, which leads to $n_{\text{th}}^{\pm}=k_{B}T\Gamma(\Omega)\Omega/\hbar\Gamma_{m}(\omega_{m}^{\pm})^{2}$.
    We also assumed that the environmental temperature $T$, 
    effective mechanical decay rate $\gamma_{m}$,
    mechanical frequency $\Omega$, and optical decay rate $\kappa_{\pm}$
    are fixed, which are given in Table \ref{tab:parameter}.
    The entanglement generation between two oscillators is achieved for the region
    in dark brown $\epsilon_{\rm cr}>0$ in the left and right panels, including black circles.
    We also show the entanglement behavior of each point as a function $\zeta$ in Figure \ref{fig:ZETA} with the same color curve.
    Lower panels: The lower left and right panels show $\omega_m^+$ and $\omega_m^-$ as functions of $\delta_-$, respectively.
    \label{fig:XEN}}
\end{figure}
\begin{figure}[t]
    \centering
\vspace{0.5cm}
    \includegraphics[width=8.2cm]{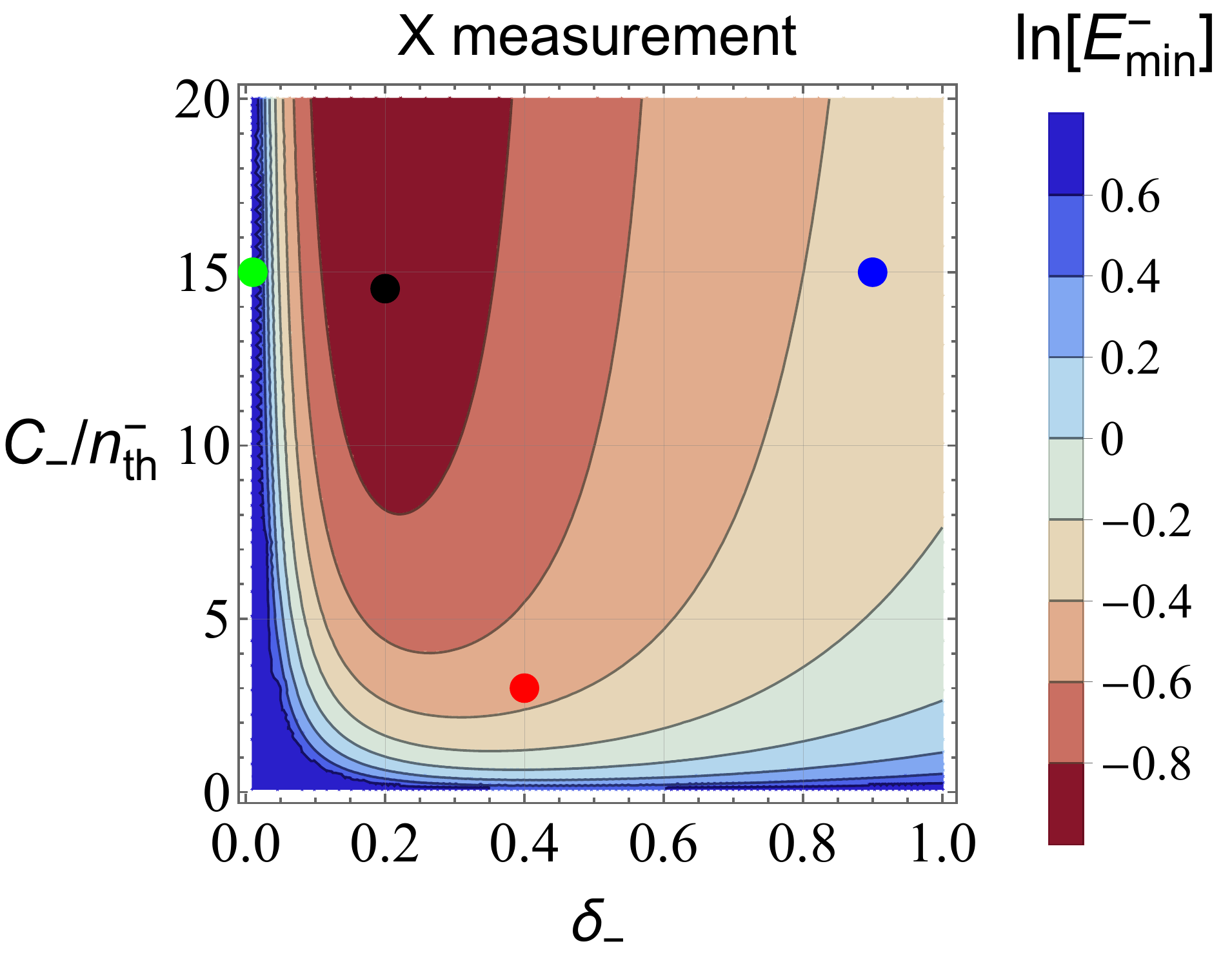}
    \hspace{1.0cm}
    \includegraphics[width=8.2cm]{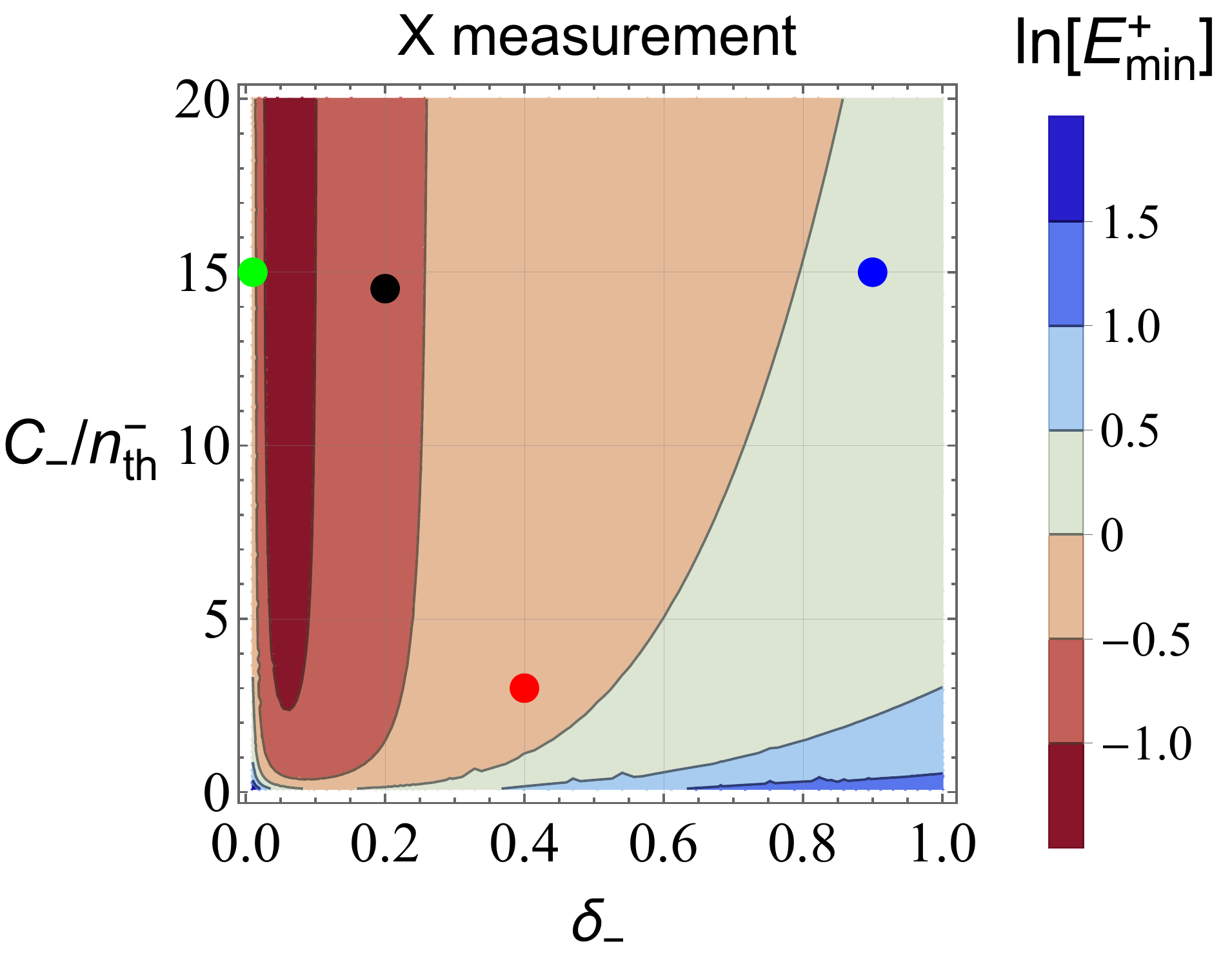}
    \caption{
    The natural logarithm of the minimum eigenvalue for the differential mode covariance matrix of the mechanical mirrors $E_{\text{min}}^{-}$ is shown 
    in the upper panel, while the same operation for the mechanical common mode $E_{\text{min}}^{+}$ is shown in the lower panel.
    Here, the covariance matrix is normalized at the frequency $\omega_{m}^{\pm}$ and each component is given by \eqref{cvcd}. (also see Appendix C)
    When $E_{\text{min}}^{\pm}<1$, the squeezed uncertainty is less than the vacuum fluctuation. 
    }\label{fig:XSQ}
\end{figure}

Next, we discuss the entanglement behavior of our results. We adopt the parameters in Table \ref{tab:parameter}, 
some of which have already been achieved \cite{Matsumoto20,MY}, whereas others are conservative parameters expected in the mid-term future.
Here we assume the structural damping $\Gamma(\omega_{m}^{\pm})=\Gamma(\Omega)\Omega/\omega_{m}^{\pm}$, which leads to
\begin{eqnarray}
    n_{\text{th}}^{\pm}=
    \frac{k_{B}T\Gamma(\Omega)\Omega}{\hbar\gamma_{m}(\omega_{m}^{\pm})^{2}}.
    \label{nth+}
    \label{nplus}
\end{eqnarray}
The environmental temperature is effectively lowered by the feedback control as $T \rightarrow T_{\rm eff}= T\Gamma/\gamma_m$, which reduces the thermal noise as in Eq.~(\ref{nplus}).
\newpage

We now consider the tabletop experiments with $\text{mg}$ scale mirrors, so assume that the mechanical frequency $\Omega$, 
effective mechanical decay rate $\gamma_{m}$,
bath temperature $T$, optical decay rate $\kappa_{\pm}$, and ratio of the optical decay rate $\zeta$ are fixed as those in Table \ref{tab:parameter}; the
variable parameters are the 
bare mechanical decay rate $\Gamma$
and the detuning $\Delta$.
In these parameter regions, the measurement of the optical phase quadrature is experimentally difficult.
Thus, we primarily consider the measurement of both the optical amplitude quadratures of the common mode $X_{+}$ and the differential mode $X_{-}$.

The upper left panel of 
Figure \ref{fig:XEN} plots $\epsilon_{\rm cr}$ for 
the $X$ measurement of both the optical amplitude quadratures $X_{\pm}$ 
as a function of 
the quantum cooperativity $C_{-}/n_{\text{th}}^{-}$ and the normalized detuning $\delta_{-}$.
Here, the quantum cooperativity depends on both $\Gamma$ and $\delta_{\pm}$, and
the upper right panel of Fig.~\ref{fig:XEN} shows the same plot as a function of $\Gamma$ and $\delta_{-}$.
Entanglement appears for $\epsilon_{\rm cr}>0$, 
which is achieved for
the dark brown regions in the upper left and right panels,
including the black circle.
The minimum quantum cooperativity required to generate the entanglement is 
$C_{-}/n_{\text{th}}^{-}\simeq3$,
and $\delta_-\simeq0.1\sim0.2$ is advantageous in generating entanglement for the $X$ measurement.
The lower panels of Fig.~\ref{fig:XEN} show the behavior of the frequency $\omega_{m}^{\pm}$ as a function of the detuning $\delta_{-}$.
In the $X$ measurement, the entanglement is optimized near the peak of the frequency of both modes,
$800\text{[Hz]}\simlt\omega_{m}^{+}/2\pi\simlt1.4\text{[kHz]}$ and
$400\text{[Hz]}\simlt\omega_{m}^{-}/2\pi\simlt500\text{[Hz]}$.

\begin{figure}[H]
    \begin{minipage}{1\linewidth}
    \centering
    \subcaption{GREEN}
    \includegraphics[width=0.30\linewidth]{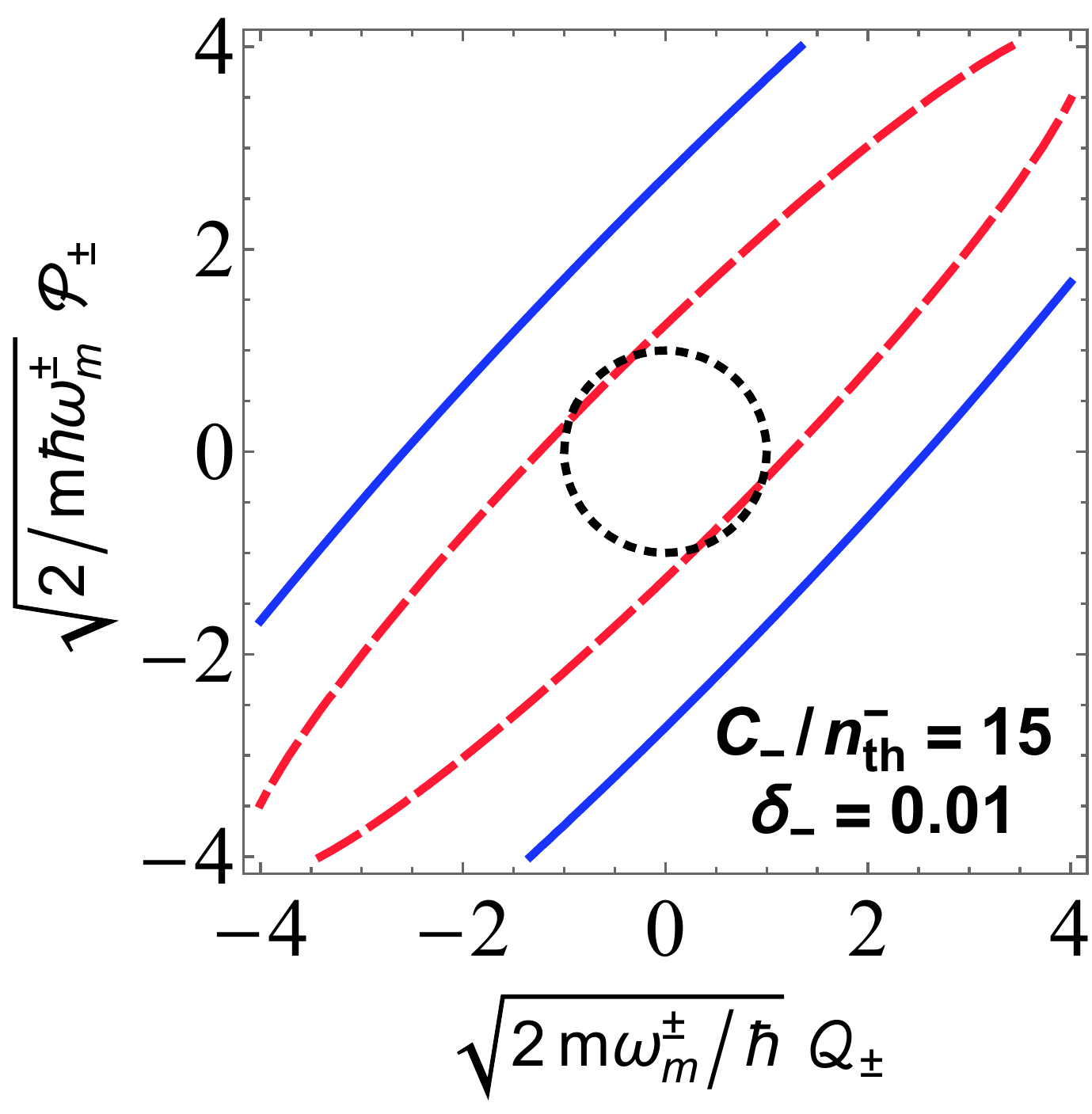}
    \hspace{1cm}
    \includegraphics[width=0.30\linewidth]{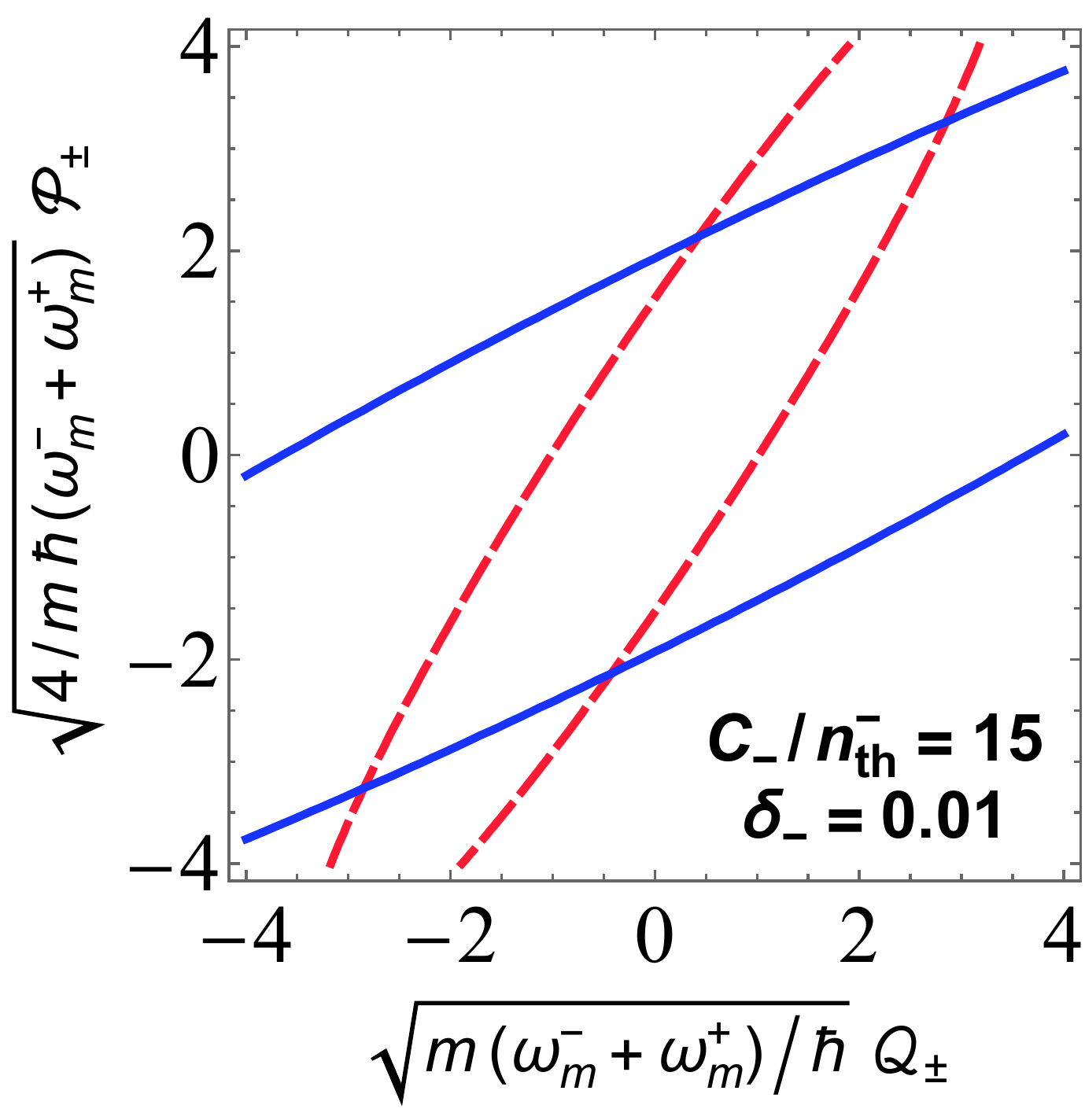}
    \end{minipage}\\
    \begin{minipage}{1\linewidth}
    \centering
    \subcaption{BLUE}
    \includegraphics[width=0.31\linewidth]{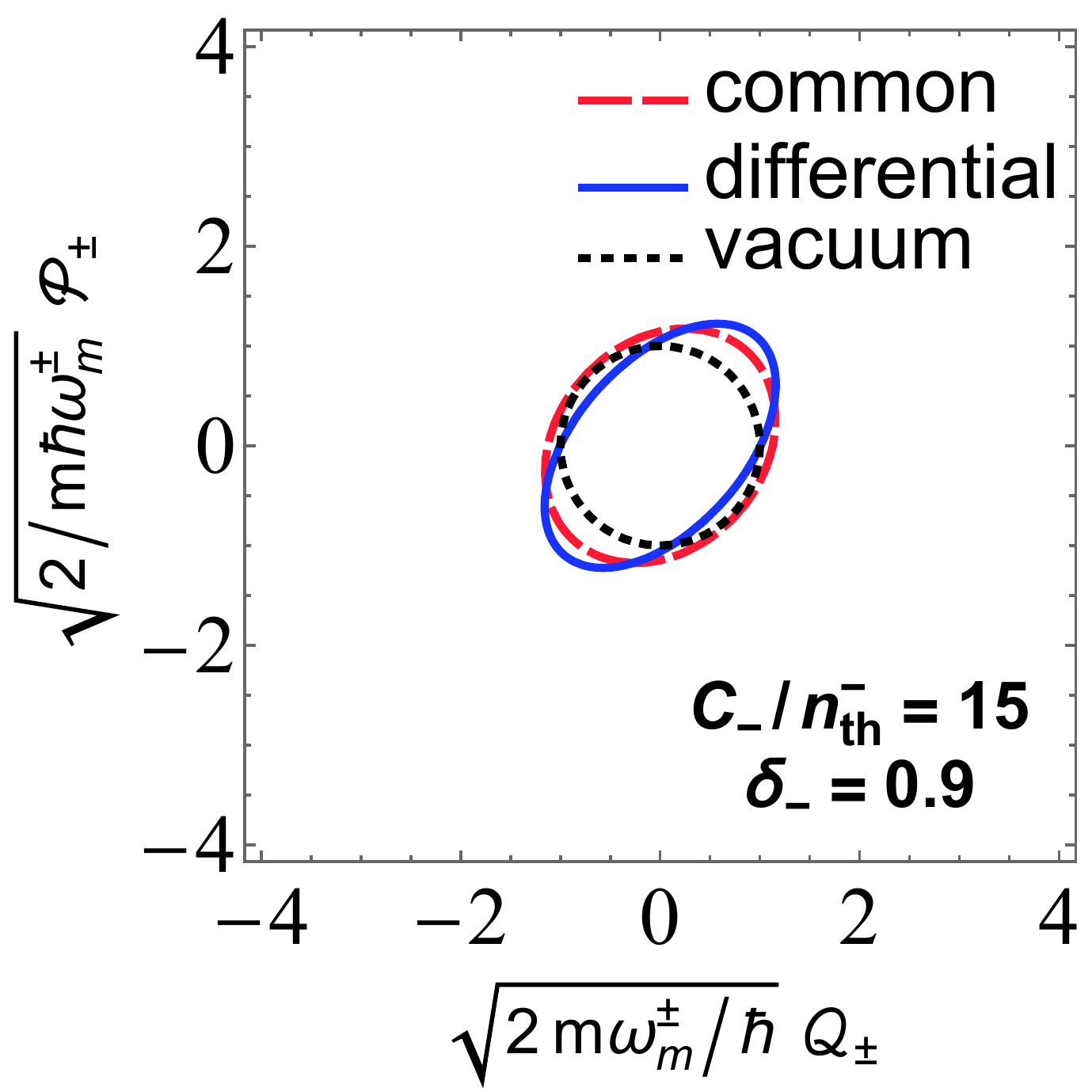}
    \hspace{1cm}
    \includegraphics[width=0.30\linewidth]{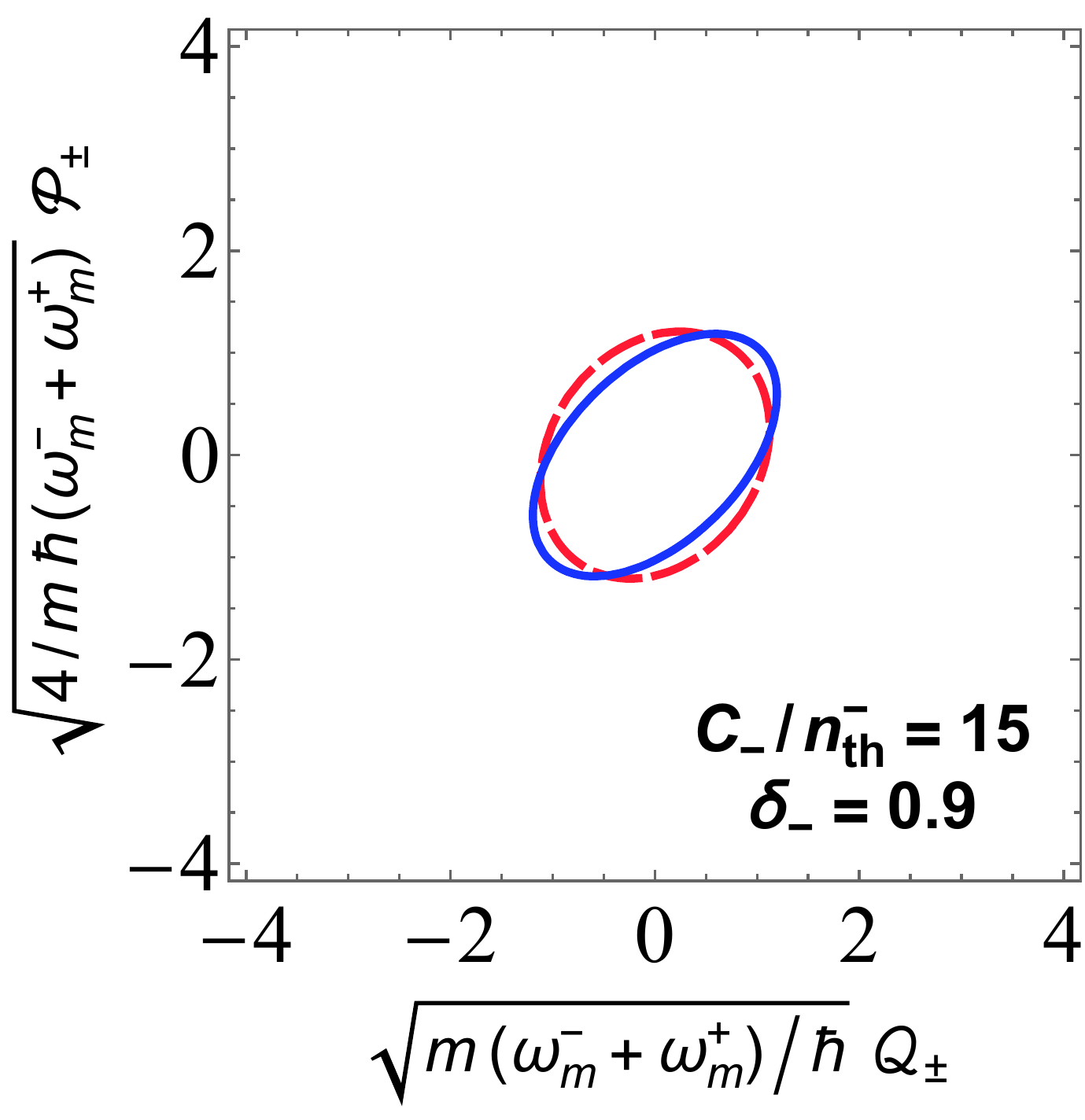}
    \end{minipage}\\
    \begin{minipage}{1\linewidth}
    \centering
    \subcaption{RED}
    \includegraphics[width=0.30\linewidth]{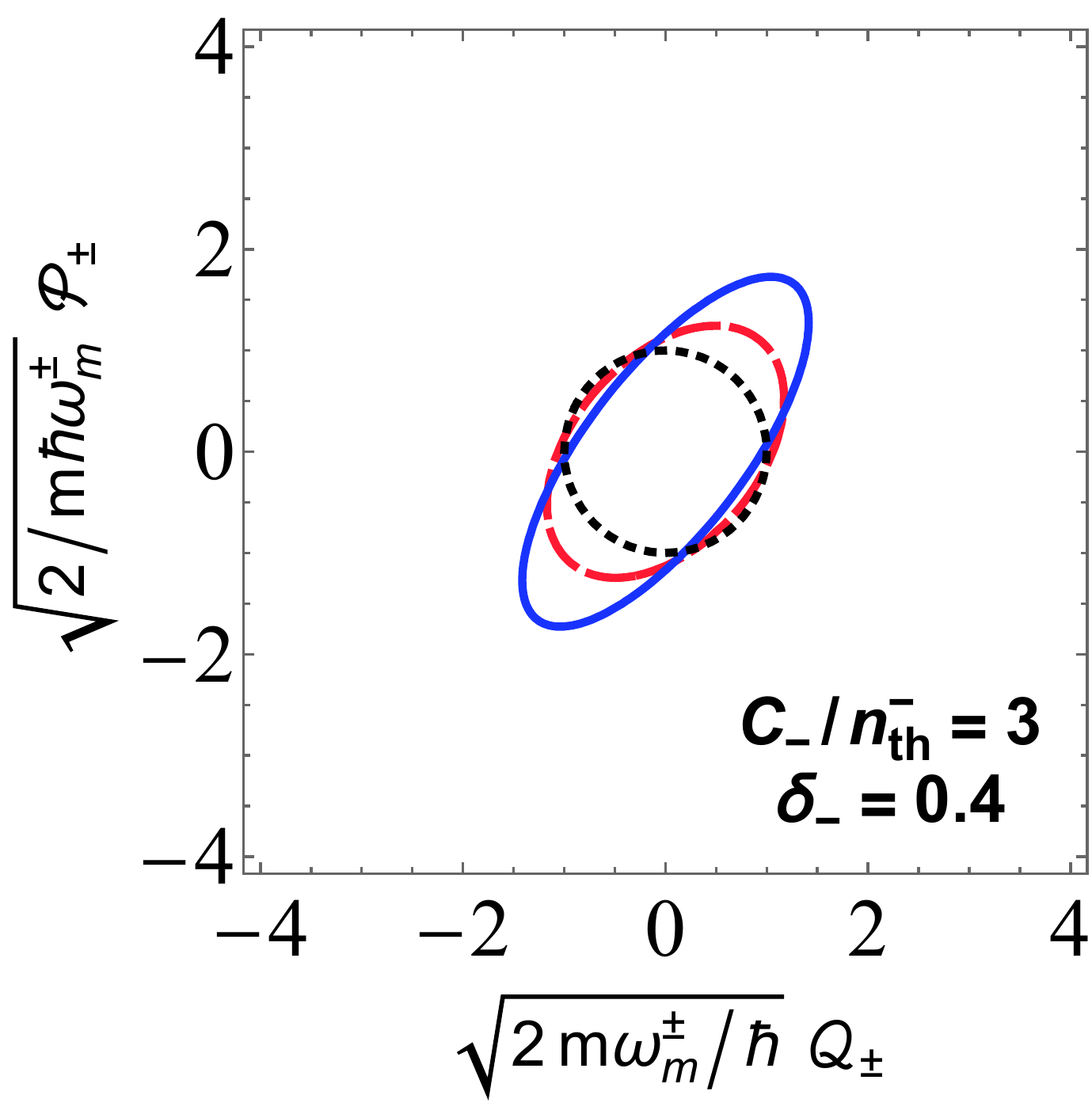}
    \hspace{1cm}
    \includegraphics[width=0.30\linewidth]{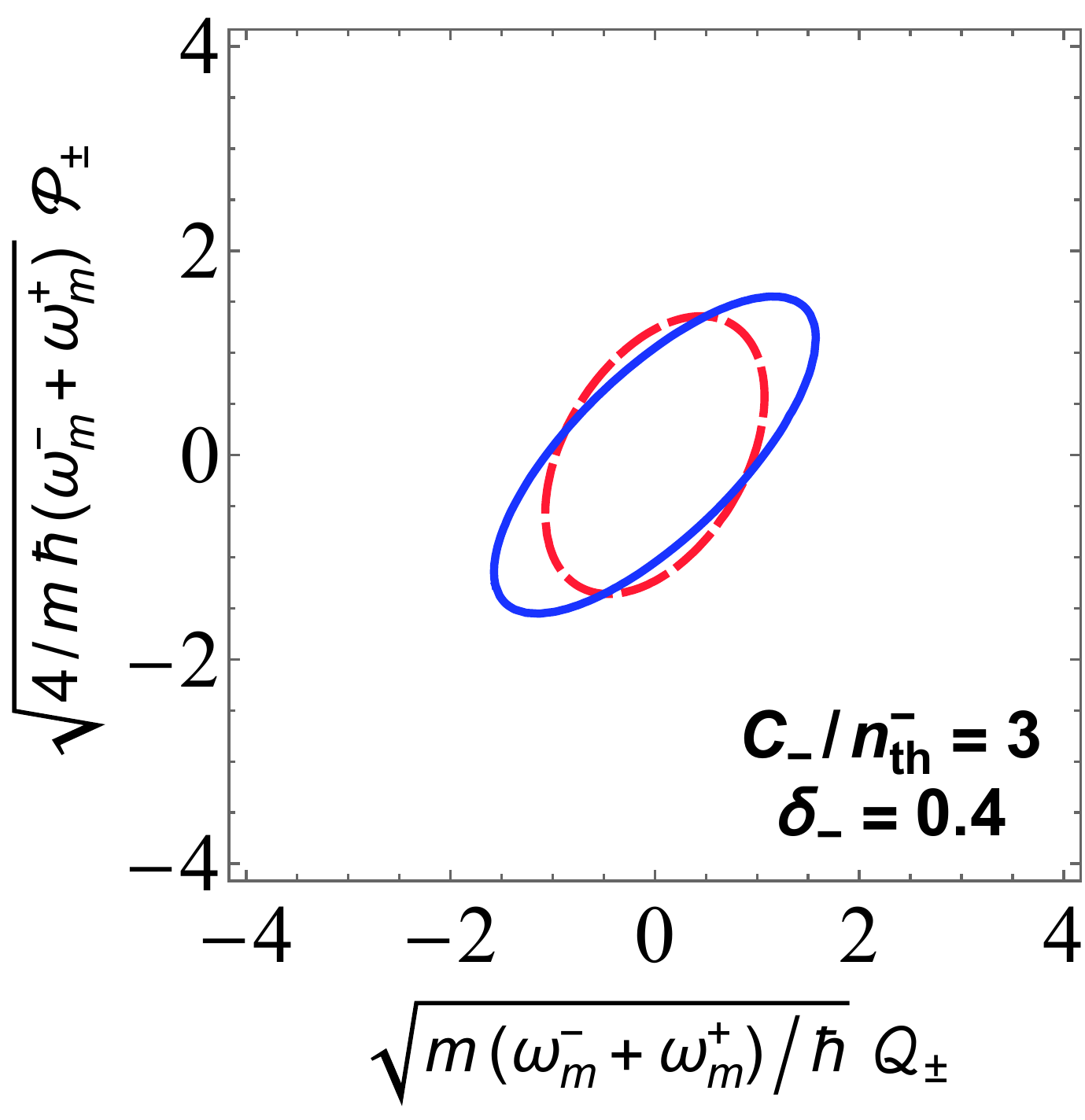}
    \end{minipage}\\
    \begin{minipage}{1\linewidth}
    \centering
    \subcaption{BLACK}
    \includegraphics[width=0.30\linewidth]{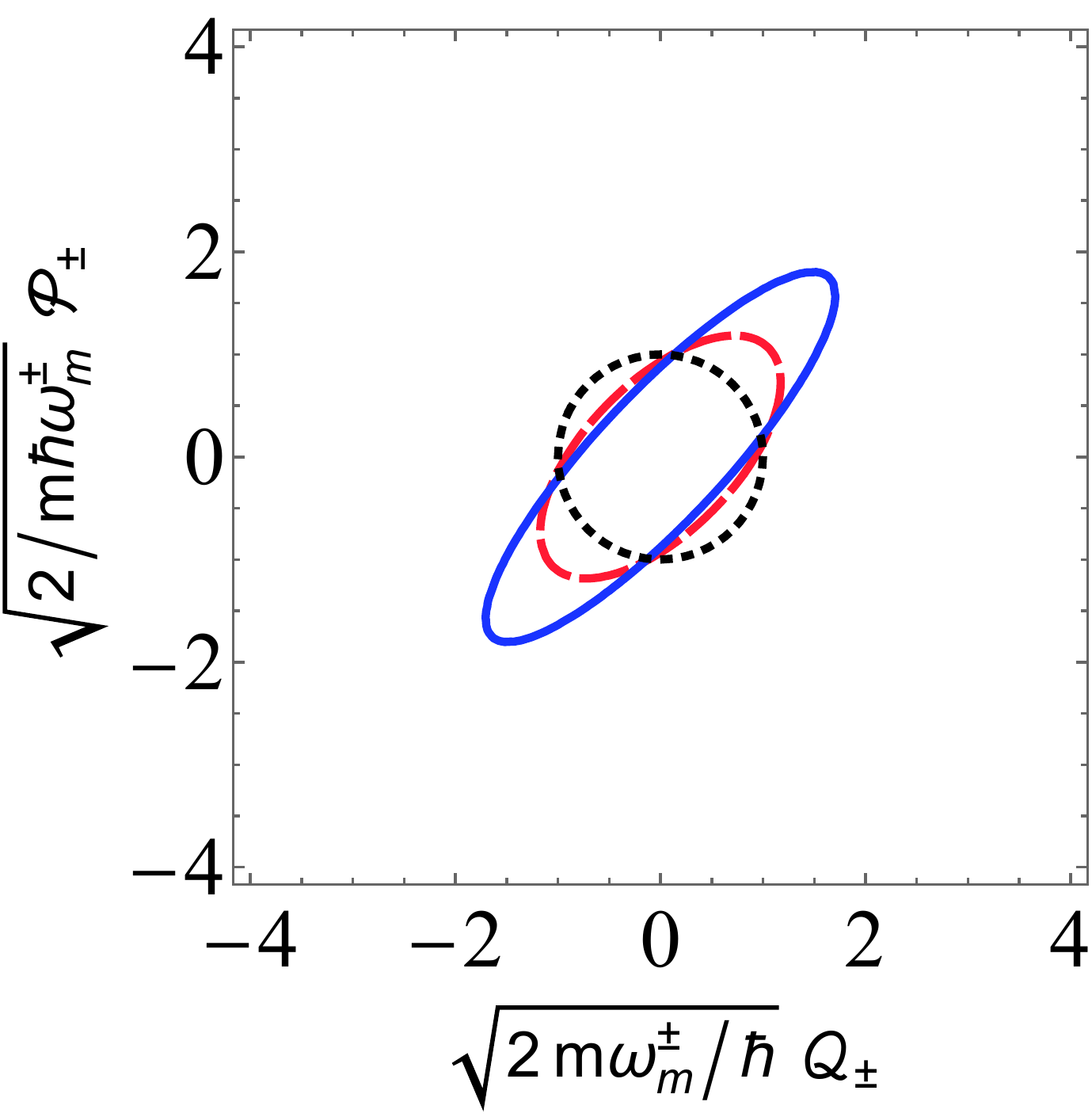}
    \hspace{1cm}
    \includegraphics[width=0.30\linewidth]{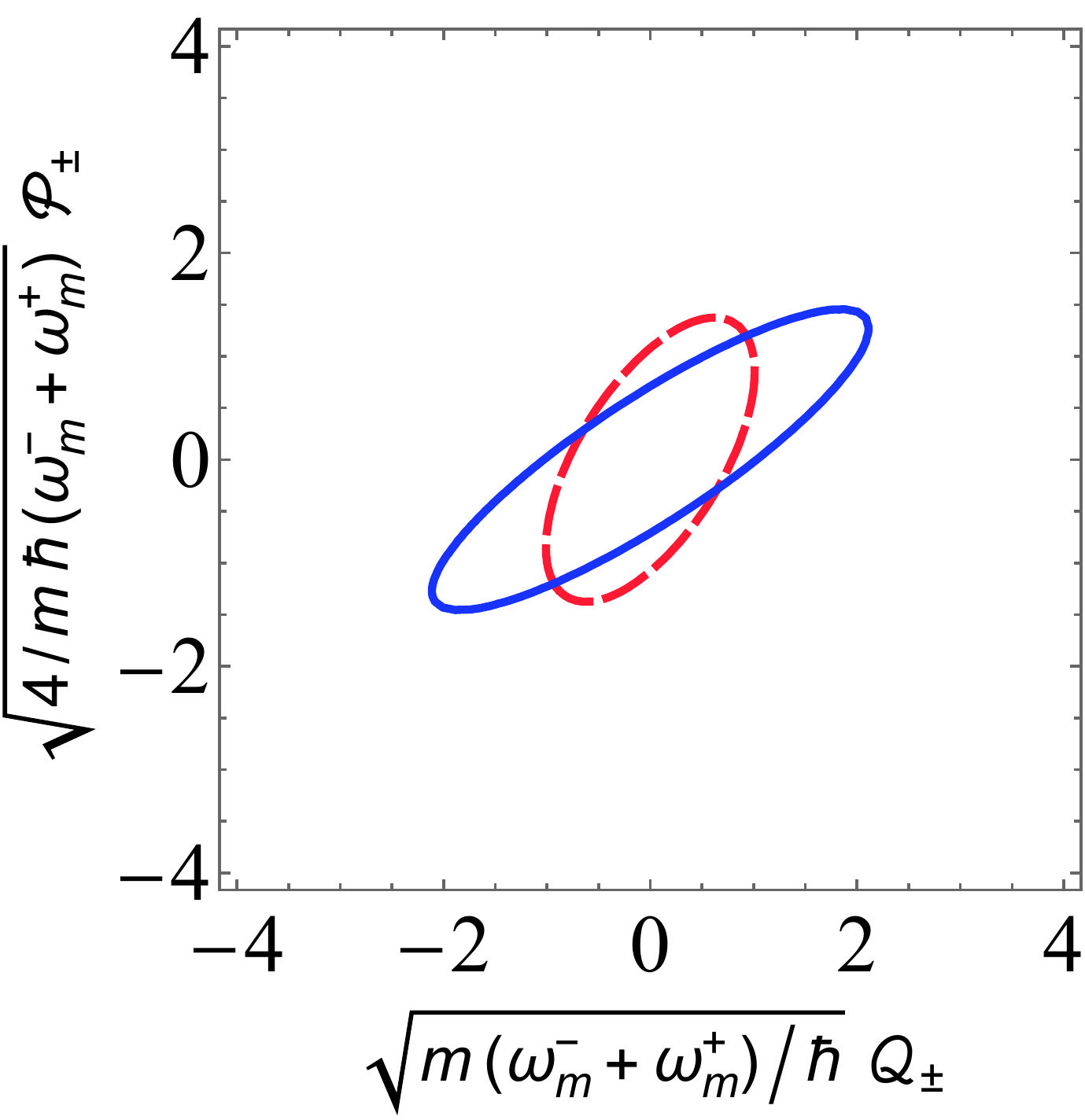}
    \end{minipage}
    \caption{
    The Wigner ellipses of the mechanical common mode (red dashed line), mechanical differential mode (blue solid line), and ground state (black dotted line).
    The covariance matrices of common and differential modes' mechanical mirrors are normalized with the frequency $\omega_{m}^{\pm}$ in the left panels and with $(\omega_{m}^{+}+\omega_{m}^{-})/2$ in the right panels.
    Each panel corresponds to the parameters specified by the colored
    circle in Fig.~\ref{fig:XEN}: (a) green circle, (b) blue circle, (c) red circle, and (d) black circle.
    The circles with the same color in Figs.~2, 3, 5, 6, and 7 assume the same parameters.
    }
    \label{fig:XPH}
\end{figure}

The quantum-squeezed state does not always mean the entanglement between the mirrors. Namely, there is no one-to-one connection between the quantum-squeezed state and the entanglement, which is demonstrated in Figures \ref{fig:XSQ} and \ref{fig:XPH}.
Figure \ref{fig:XSQ} shows the natural logarithm of the minimum eigenvalue for the mechanical covariance matrix \eqref{cvcd}, for 
$X$ measurement as a function of $C_{-}/n_{\text{th}}^{-}$ and $\delta_{-}$. 
When the minimum eigenvalue is less than $1$, the state is quantum squeezed because the squeezed uncertainty is less than that of the vacuum state.
Therefore, the region, $C_{-}/n_{\text{th}}^{-}\simgt1$
and $0.1\simlt\delta_-\simlt0.8$ roughly,
satisfies the condition of a quantum-squeezed state.

Fig. \ref{fig:XPH} shows the curves as the contour for the Wigner function satisfying $W=e^{-1}W_\text{max}$.
The red dashed, blue solid, and black dotted curves correspond to the Wigner contours of the mechanical common mode, the mechanical differential mode, and the ground state.
Each panel assumes that the parameters
$C_-/n_{\text{th}}^{-}$ and $\delta_{-}$ correspond to the colored circles in Figs. \ref{fig:XEN} and \ref{fig:XSQ}, respectively.
Panel (a) in Fig.~\ref{fig:XPH}, which corresponds to the green circles
in Figs.~\ref{fig:XEN} and \ref{fig:XSQ}, respectively, shows the case when neither mode is quantum squeezed
and the entanglement does not occur.
Panel (b) and (c), corresponding to the blue and red circles,
illustrate the cases in which only mechanical differential modes are quantum squeezed and both the mechanical common and differential modes are quantum squeezed, respectively.
However, the entanglement is not generated in the case for the panel (b) and (c). 
Thus, the behavior of (b) and (c) on the negativity is similar, but the behavior on the squeezing is different as is shown in Fig.~\ref{fig:XSQ}.
Panel (d), corresponding to the black circle 
represents an experimentally feasible parameter expected from the proposed technique \cite{Matsumoto20,MY} by using $C_{\pm}$, $n_{\text{th}}^{\pm}$, and $Q_{\pm}$ in Table \ref{tab:parameter}.
In this case, both the modes are quantum squeezed and the entanglement is generated.

\begin{figure}[t]
\centering
\includegraphics[width=7.5cm]{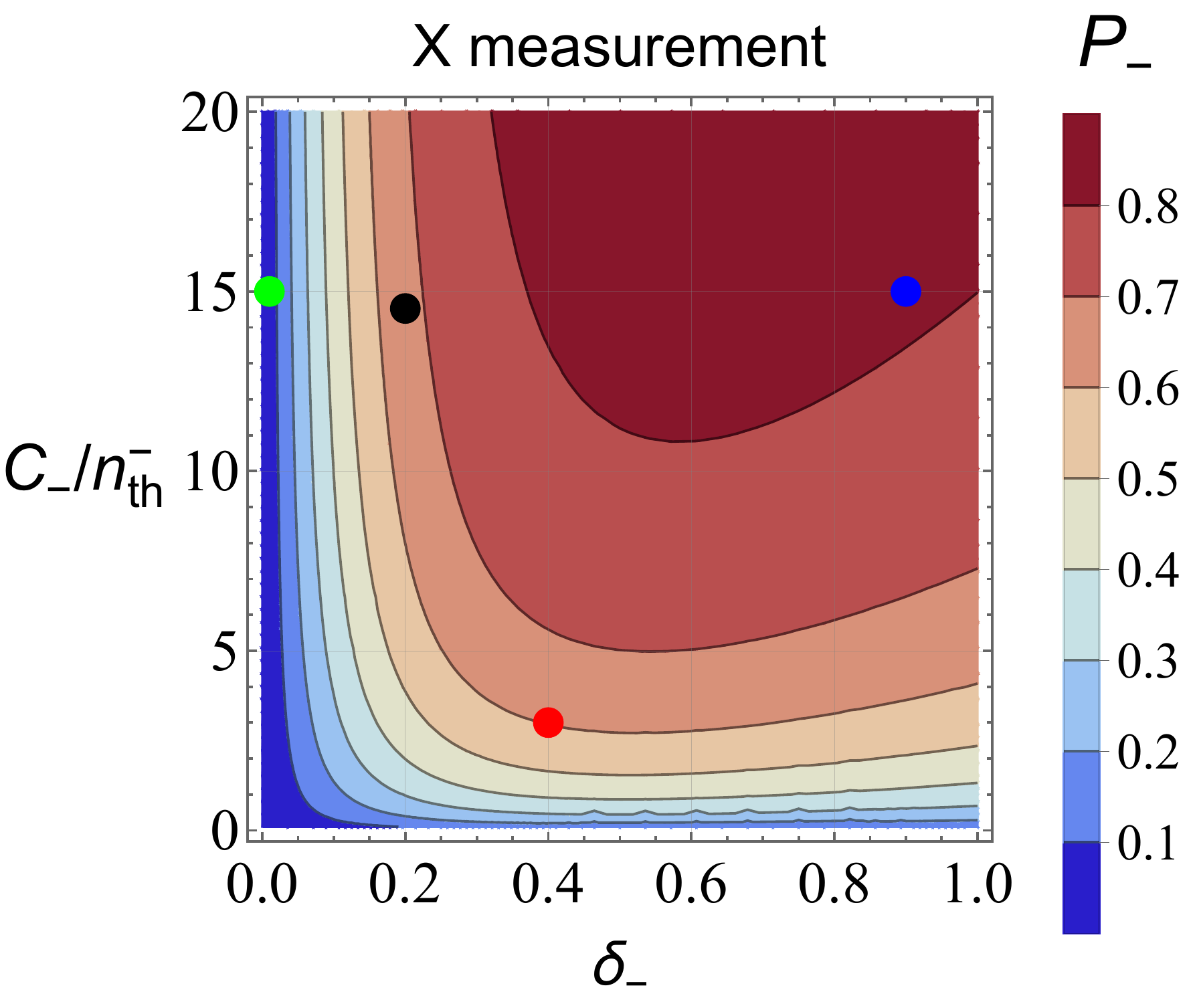}
\hspace{0.5cm}
\includegraphics[width=7.5cm]{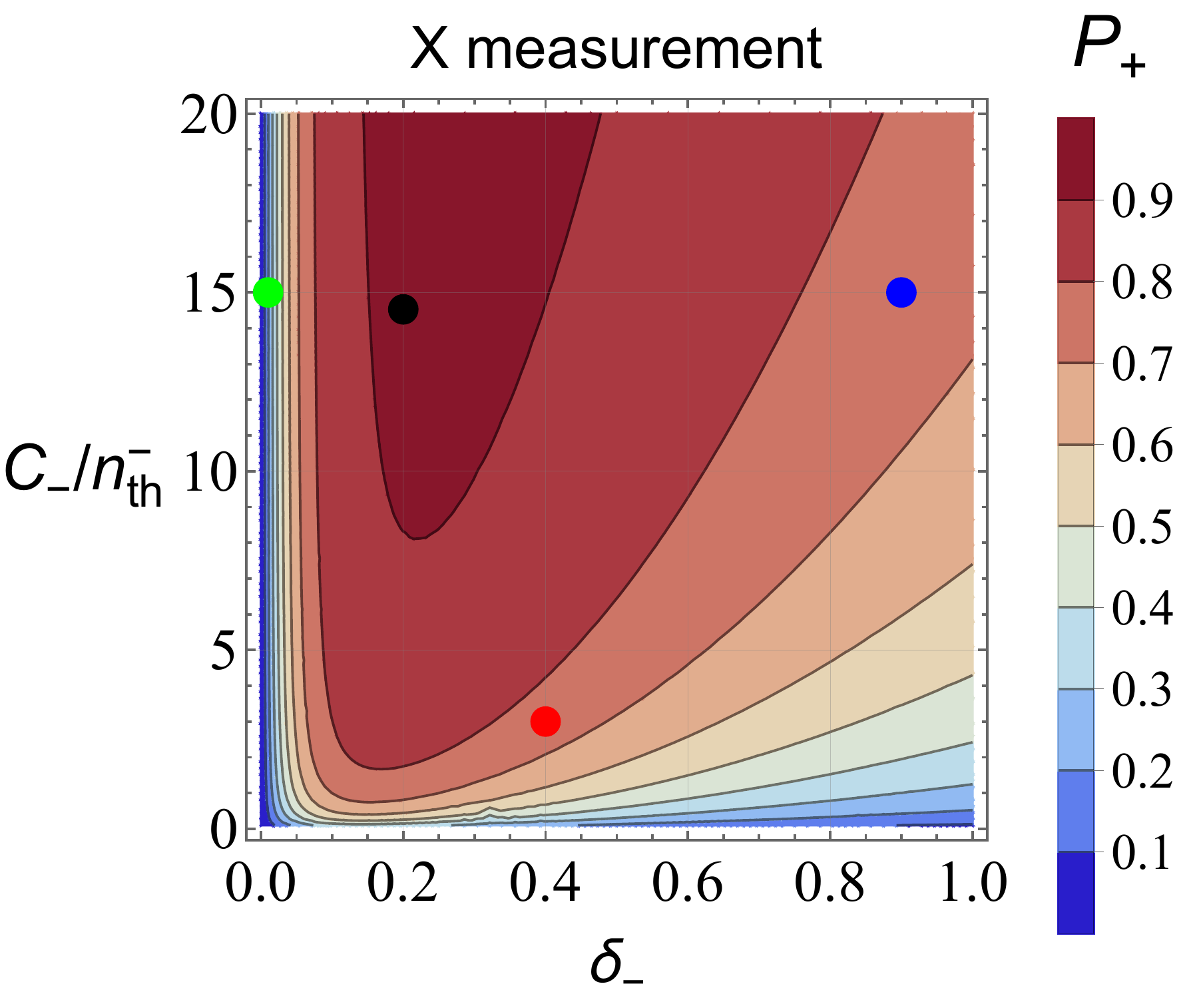}
\caption{
The behavior of the purity of 
differential (upper panel) and common (lower panel) modes.
By combining this figure and the upper panels of  Fig.~\ref{fig:XEN}, one can roughly read that $P_-\simgt0.5$ and $P_+\simgt0.8$ are necessary at $\delta_-=0.2$ for generating the entanglement.
}
\label{fig:purity}
\end{figure}

Now we discuss the relationship between quantum squeezing and entanglement.
The red circle in the upper panels of Figs.~\ref{fig:XEN} and \ref{fig:XSQ} demonstrates that squeezing does not necessarily imply the generation of entanglement. 
As shown in panel (c) of Fig.~\ref{fig:XPH}, where entanglement is not generated in Fig.~\ref{fig:XEN}, we find that the mechanical differential and common modes are in a quantum-squeezed state.
We infer that purity
plays a role in the generation of entanglement.
Figure
\ref{fig:purity} plots the purity as a function of $C_{-}/n_{\rm th}^{-}$ and $\delta_{-}$ for the 
mechanical differential mode (upper panel) and common mode (lower panel).
We find that the high purity and the quantum-squeezing are necessary for generating entanglement.
At $\delta_{-}=0.2$, we roughly need $P_{-}\simgt0.5$ and $P_{+}\simgt0.8$ for generating the entanglement between the two mirrors.
The measurement rate in the $X$ measurement, which is the coefficient of $q_{\pm}'$ in the first term on the right side of Eq.~\eqref{Xpm}, is maximized around $\delta_{-}=0.2$.
Thus, the purity required for entanglement increases as $\delta_{-}$ shifts from $0.2$.

\begin{figure}[H]
\centering
\includegraphics[width=8.2cm]{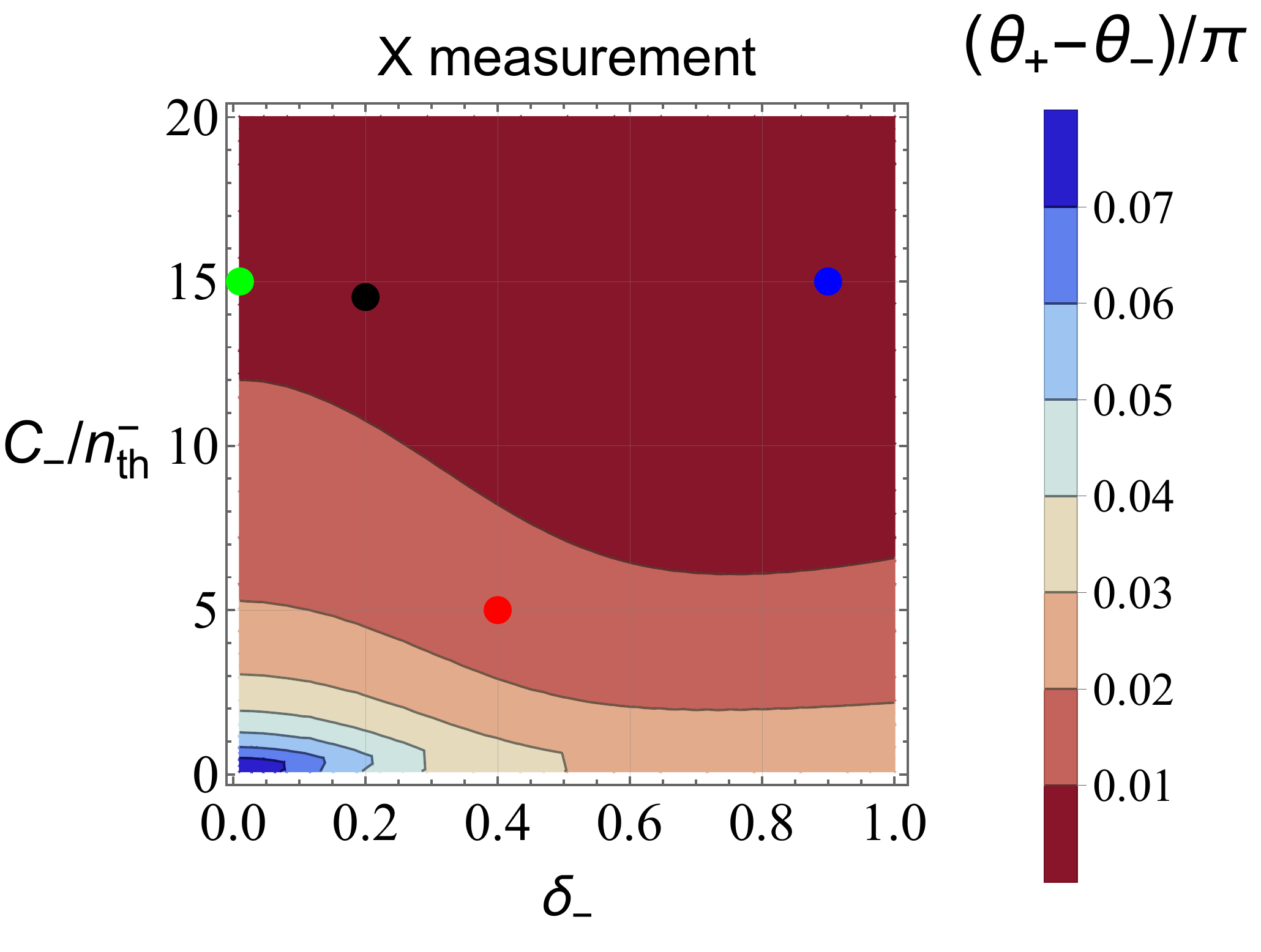}
\caption{
The behavior of the difference of two squeezing angles between the mechanical common mode and differential mode 
normalized by the frequency $\omega_{m}^{\pm}$,
which is small in our system. 
}
\label{fig:sqan}
    \centering
    \includegraphics[width=7.5cm]{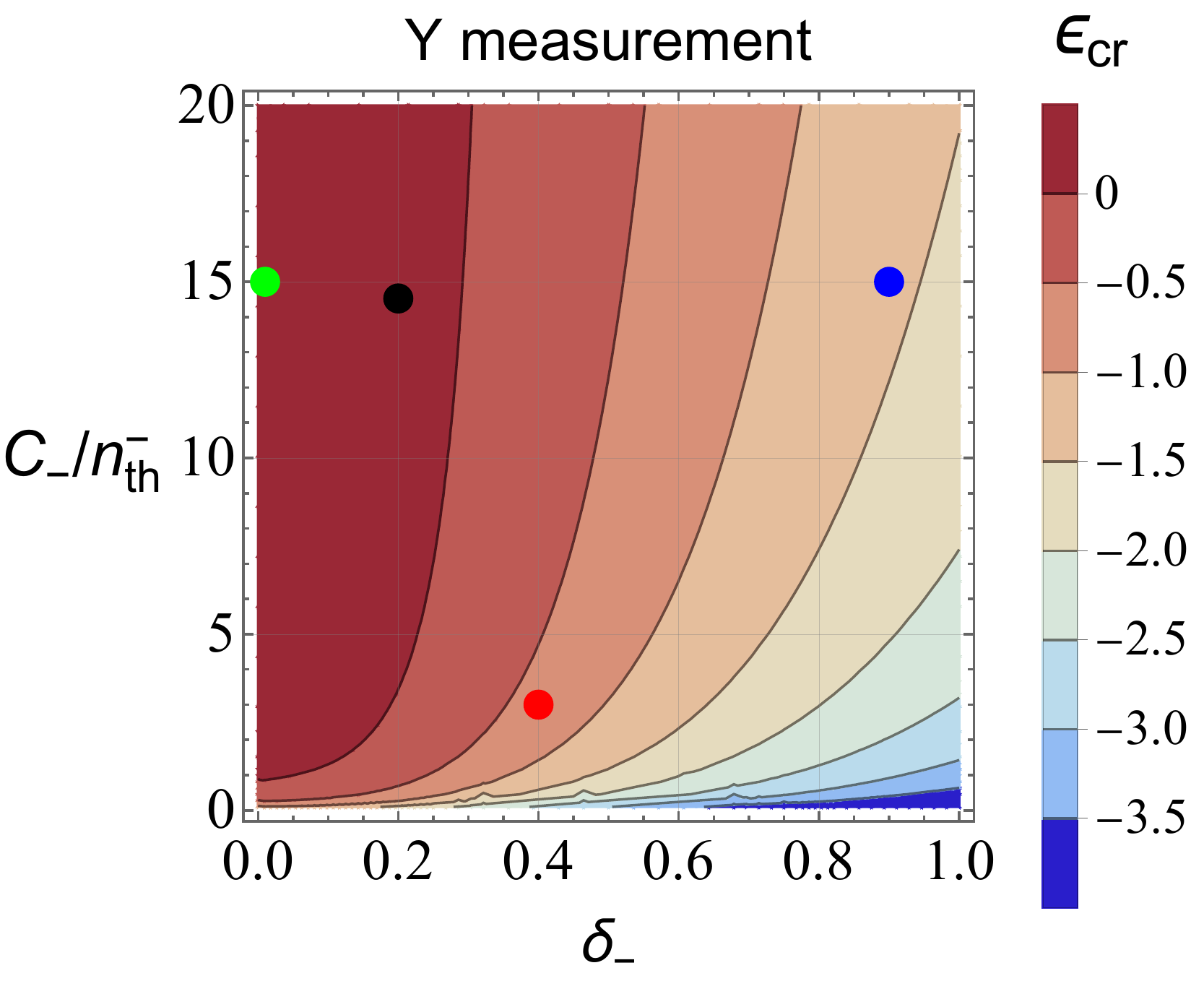}
    \hspace{1.0cm}
    \includegraphics[width=7.cm]{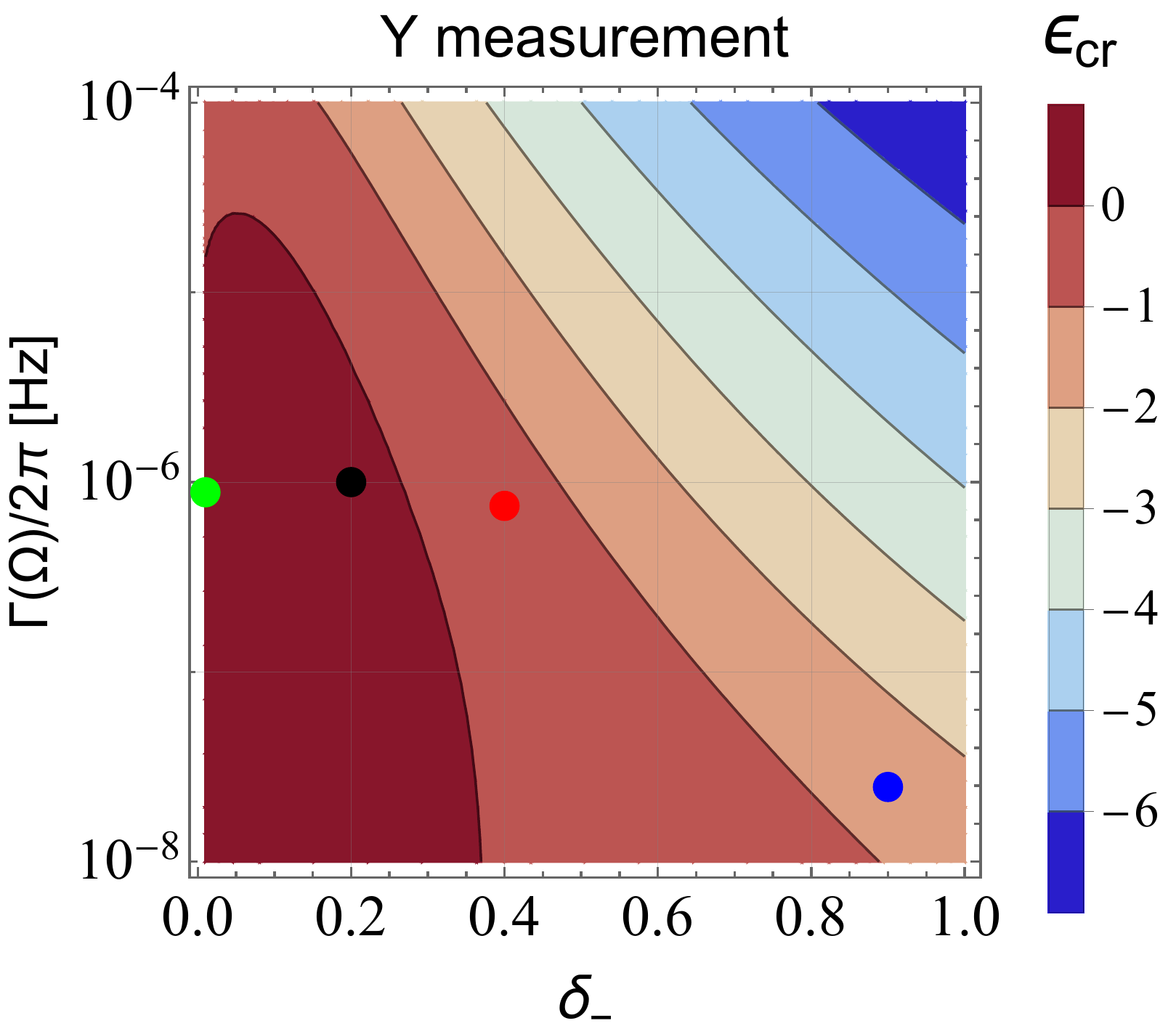}
    \caption{
    Same as the upper panels of Fig.~\ref{fig:XEN} but for the $Y$ measurement.
    We consider the $Y$ measurement of both the optical phase quadratures of the common and differential modes $Y_{\pm}$.}
    \label{fig:YEN}
\end{figure}

From an analogy of the entanglement generation by passing the two squeezed beams 
through a half-beam splitter, it is known that the difference in the squeezing angles is
an important factor.
This can be read from the right panels of Fig.~\ref{fig:XPH}. 
We note that this is the property when the
Wigner ellipse is plotted with the variables normalized 
by the frequency $(\omega_{m}^++\omega_{m}^-)/2$, 
as is done in Ref. \cite{HMullerEbhardt}.
However, the left panels of Fig.~\ref{fig:XPH} show the same plots as the right panels but with the different normalization of the variables with $\omega_{\rm m}^\pm$. 
The common mode and the differential mode are normalized 
with $\omega_{m}^+$ and $\omega_m^-$, respectively,
and then the phase diagram of the vacuum state is the circle with the unit radius. Following this
normalization of the variable, the entanglement can be generated even for the cases of the small difference of squeezing angle between the mechanical common mode and the differential mode.
Figure~\ref{fig:sqan} plots the difference of squeezing angle between the common mode and the differential mode as a function of $C_{-}/n_{\rm th}^{-}$ and $\delta_{-}$, 
where squeezing angle is defined using the Wigner ellipses normalized with the $\omega_m^\pm$
as shown in the left panels of Fig.~\ref{fig:XPH}. One can see that the difference of the 
squeezing angle is quite small in the entire 
region of the plot.
We note that these differences of the normalization do not affect the entanglement at all 
because the entanglement does not depend on the normalization of the Wigner ellipse.

\begin{figure}[H]
    \begin{minipage}{1\linewidth}
    \centering
    \subcaption{GREEN}
    \includegraphics[width=0.30\linewidth]{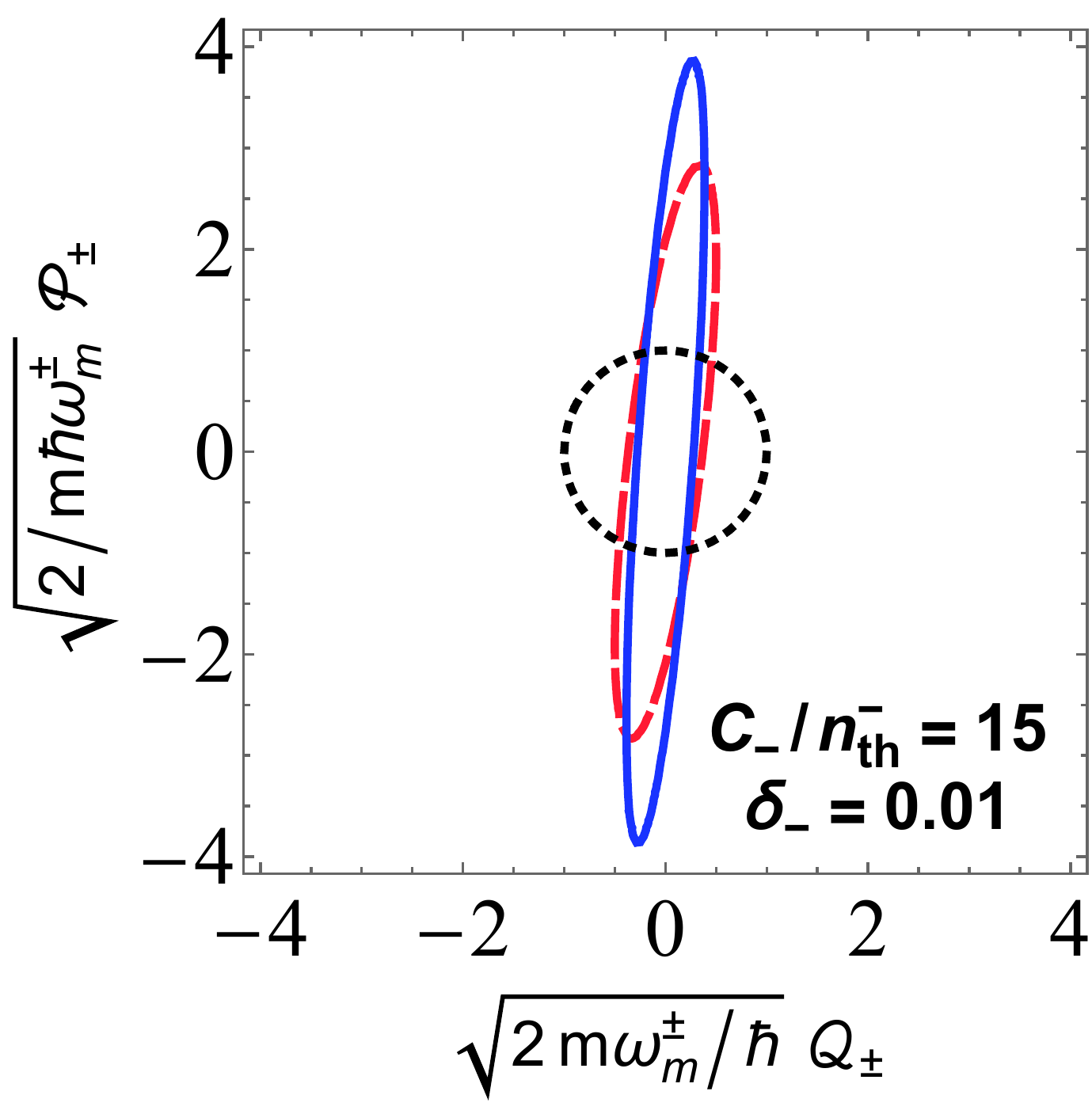}
    \hspace{1cm}
    \includegraphics[width=0.30\linewidth]{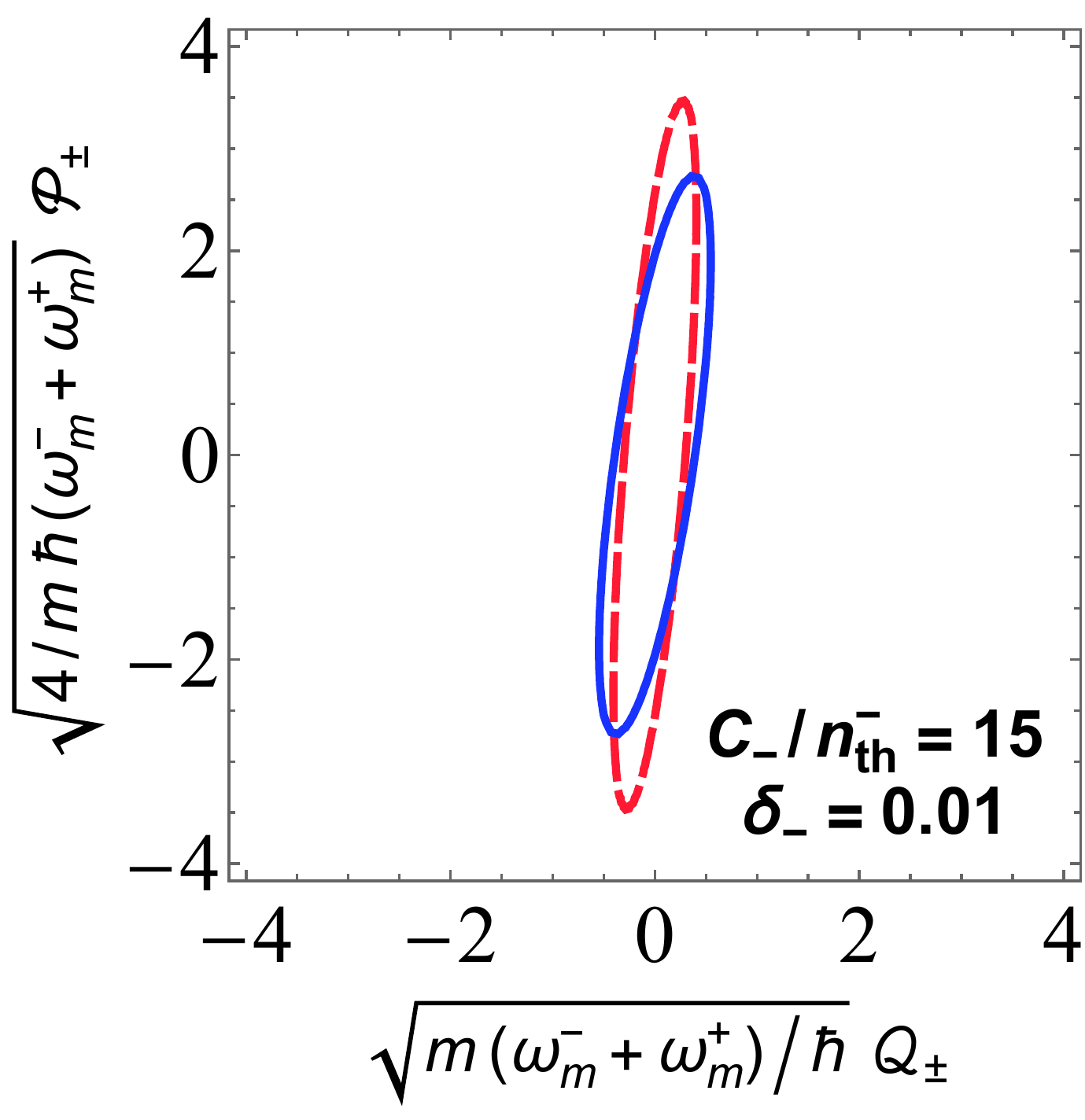}
    \end{minipage}\\
    \begin{minipage}{1\linewidth}
    \centering
    \subcaption{BLUE}
    \includegraphics[width=0.31\linewidth]{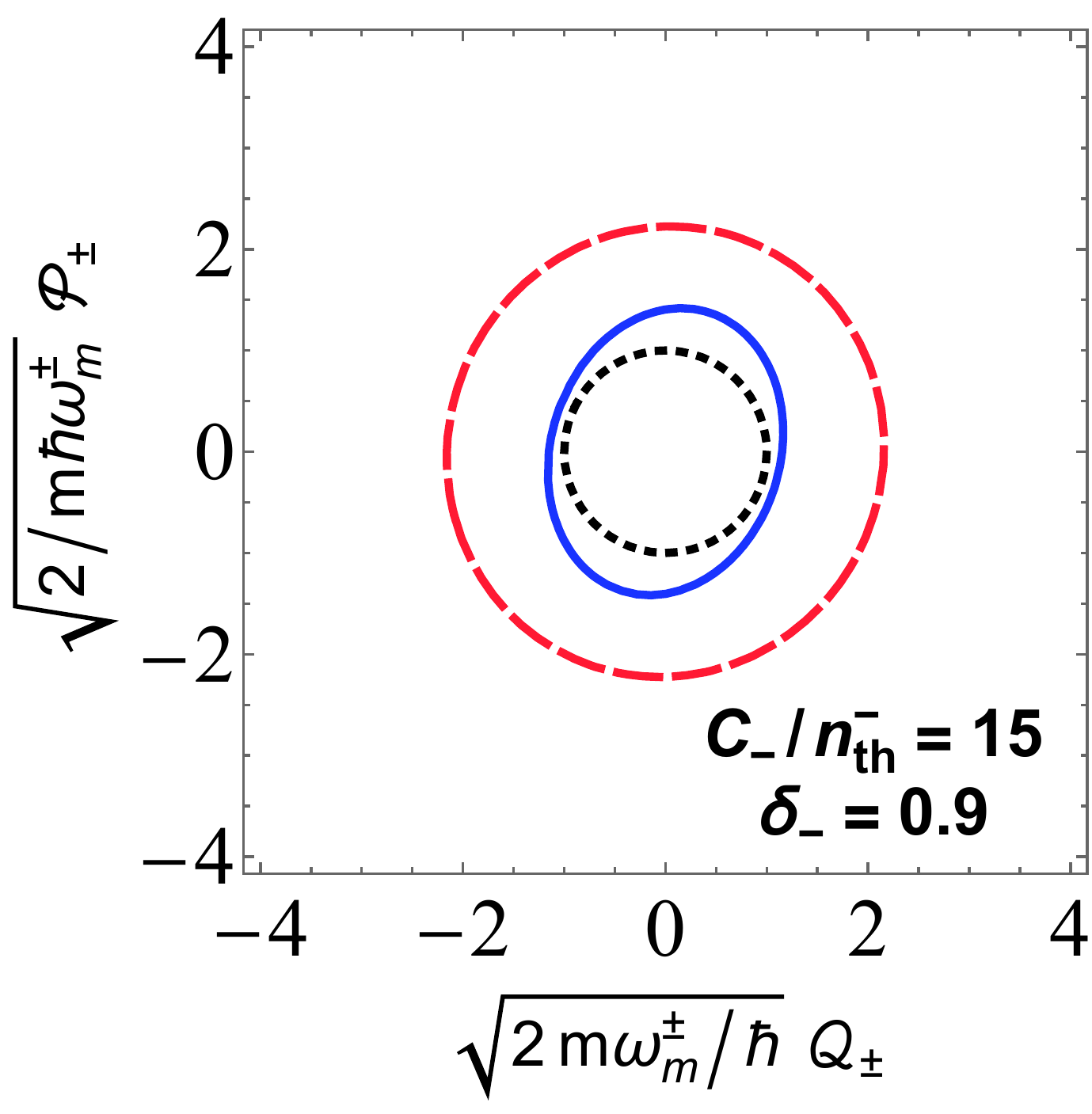}
    \hspace{1cm}
    \includegraphics[width=0.30\linewidth]{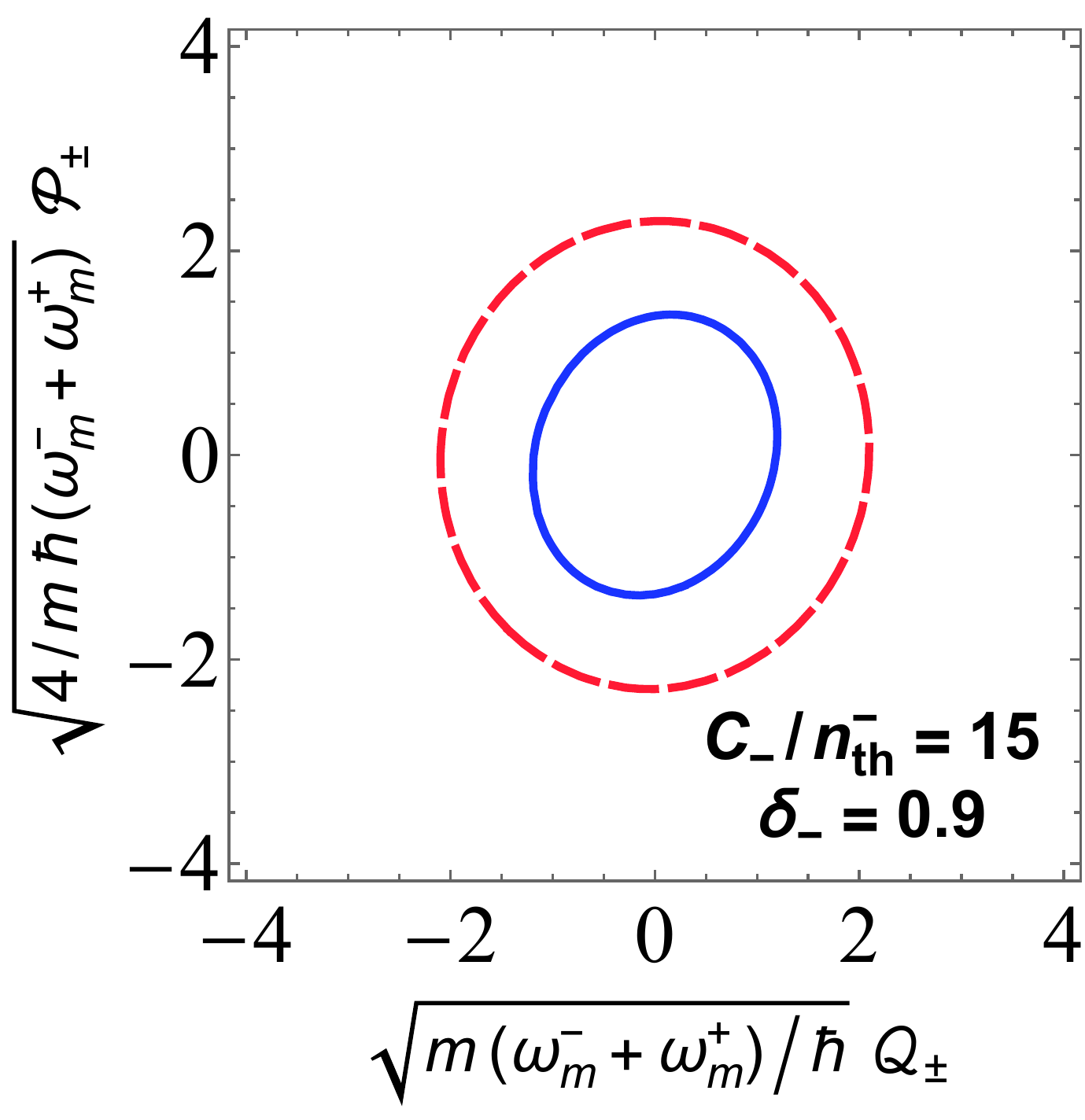}
    \end{minipage}\\
    \begin{minipage}{1\linewidth}
    \centering
    \subcaption{RED}
    \includegraphics[width=0.30\linewidth]{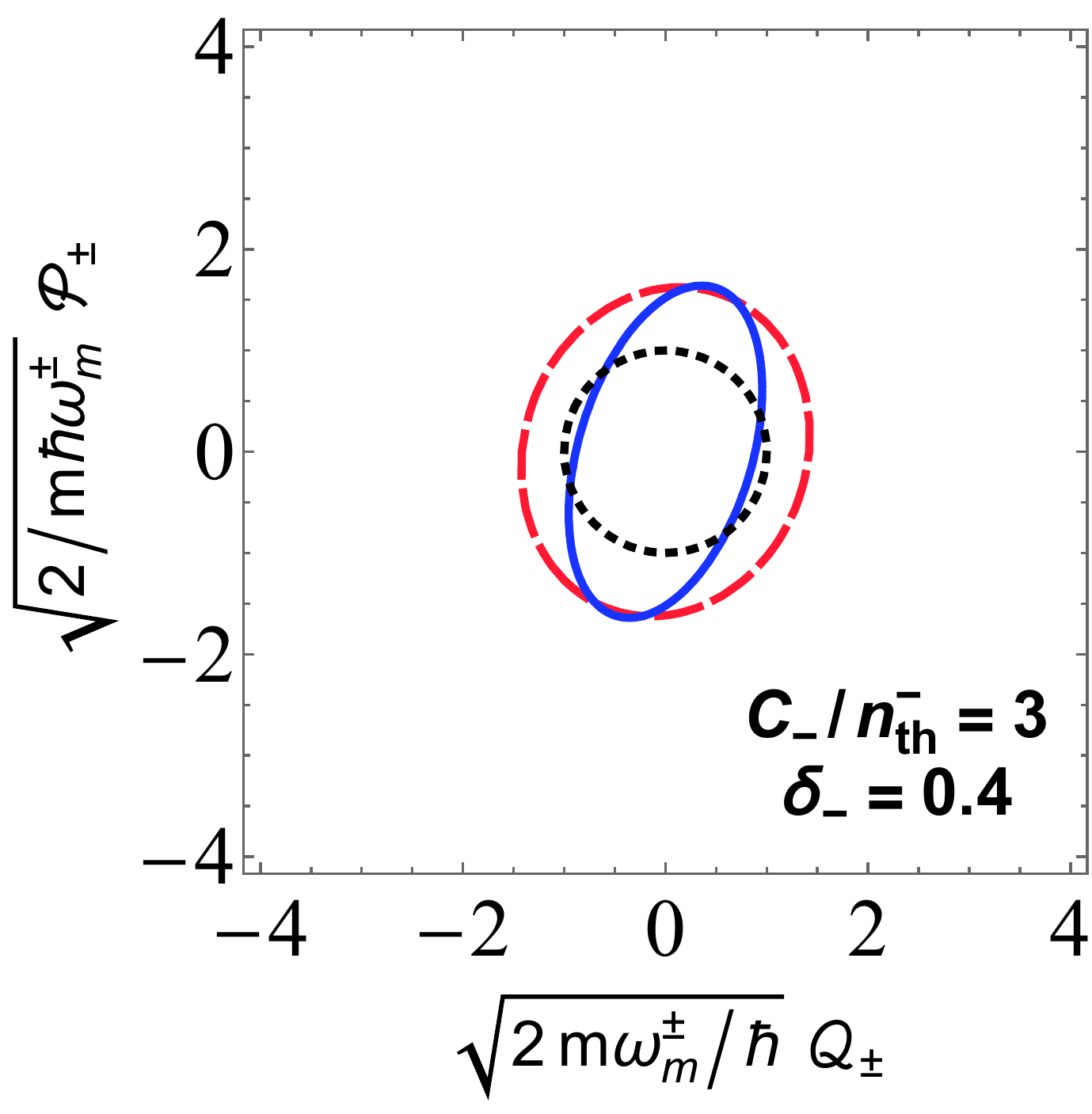}
    \hspace{1cm}
    \includegraphics[width=0.30\linewidth]{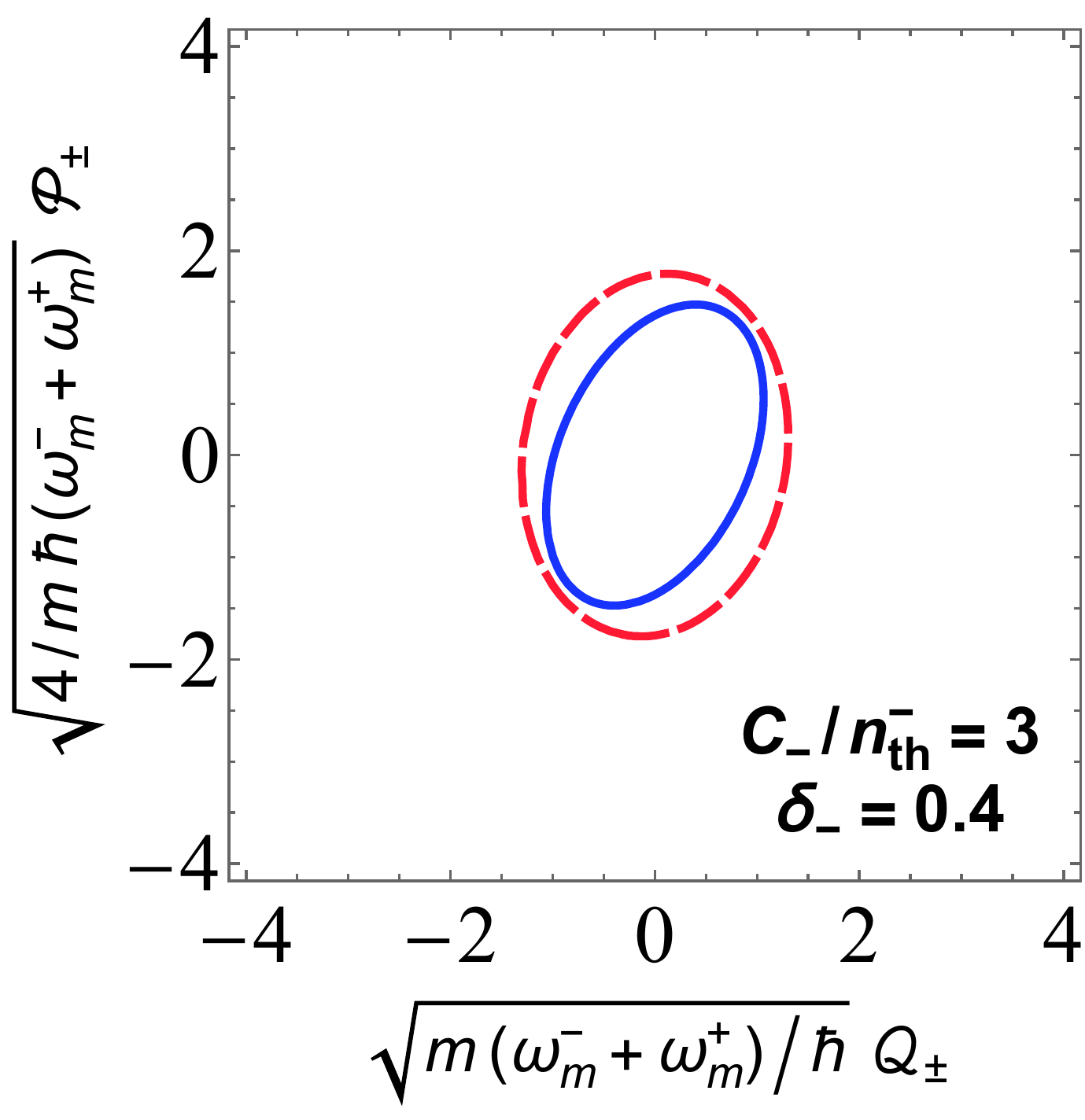}
    \end{minipage}\\
    \begin{minipage}{1\linewidth}
    \centering
    \subcaption{BLACK}
    \includegraphics[width=0.30\linewidth]{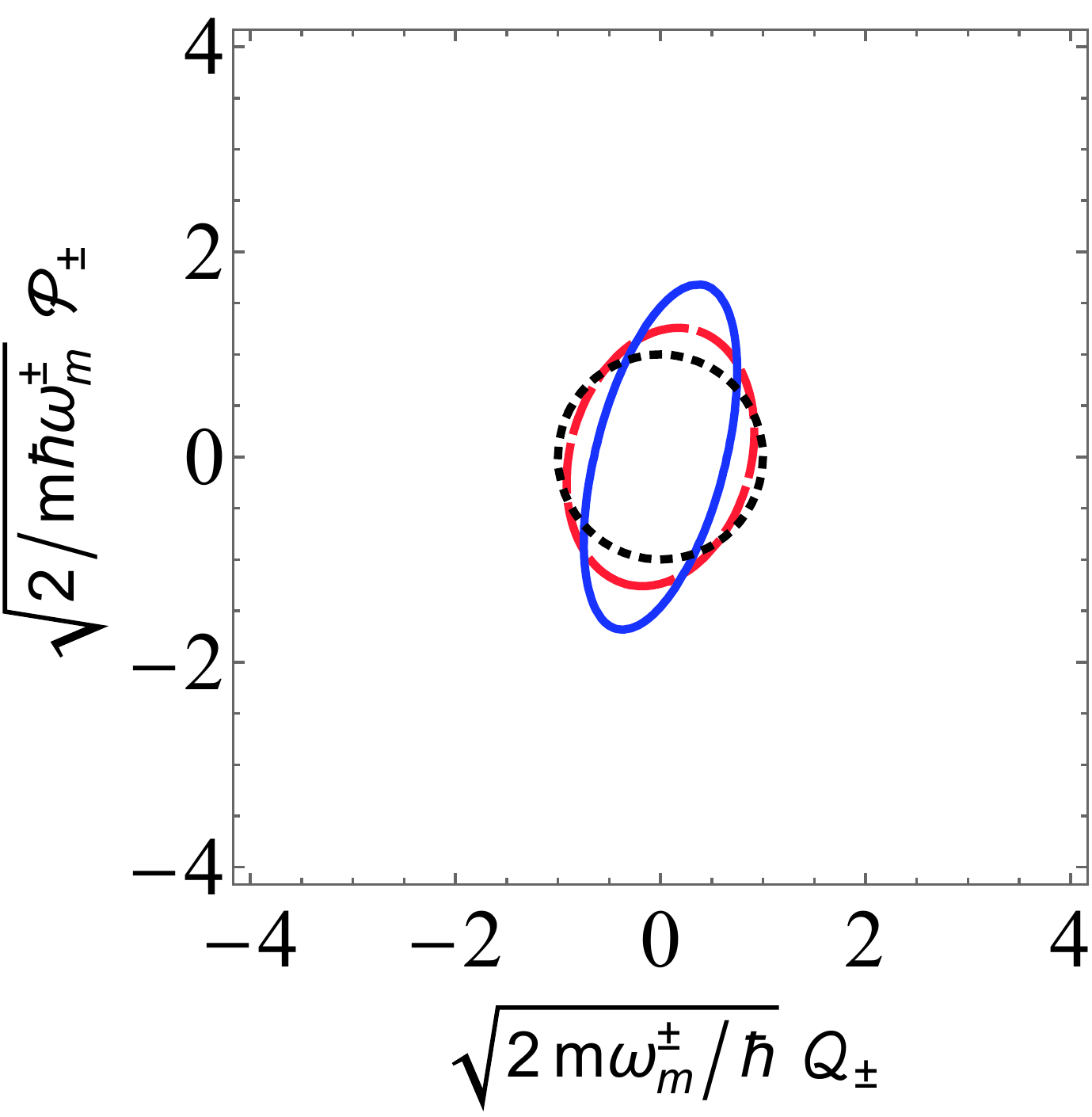}
    \hspace{1cm}
    \includegraphics[width=0.30\linewidth]{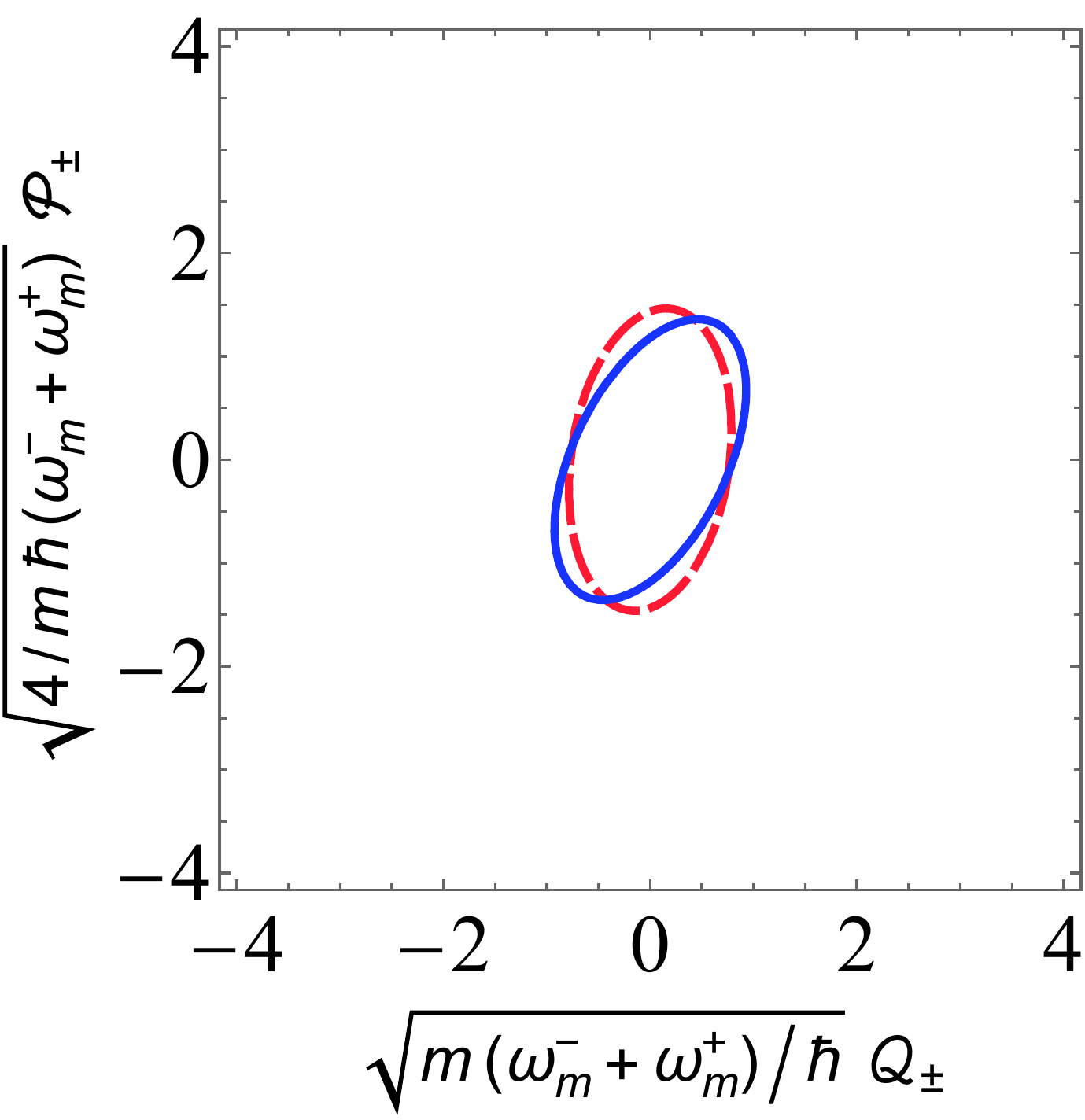}
    \end{minipage}
    \caption{
    Same as Fig.~\ref{fig:XPH} but for the $Y$ measurement, i.e., 
    the Wigner ellipses in the the phase space for the colored grid points in Fig.~\ref{fig:YEN}.
    }
    \label{fig:YPH}
\end{figure}

We next discuss the entanglement behavior and phase distribution for the $Y$ measurement
in Figs.~\ref{fig:YEN} and \ref{fig:YPH}, which are similar to those of the $X$ measurement.
Figure~\ref{fig:YEN} shows that the entanglement is more easily generated for small $\delta_{\pm}$ compared to the case in the $X$ measurement (upper panels of Fig.~\ref{fig:XEN}).
The difference is understood by the efficiency of the measurements, which is described by the first term
of the right hand side of Eqs.~(\ref{Xi}) and (\ref{Yi}).
The ellipses in Fig.~\ref{fig:YPH} show the significance of squeezing for the $Y$ measurement,
where each panel corresponds to the parameters specified by the colored circles in Fig.~\ref{fig:YEN}.
For the free-mass limit with $\delta_{\pm}=0$ and $\zeta\sim70$, the squeezing angles 
looks near orthogonal when the Wigner ellipses are normalized by the common measurement rate, which is consistent with Ref. \cite{HMullerEbhardt}.
From an experimental point of view, it should be noted that conducting the homodyne ($Y$) measurement assumed in the Table I is not easy due to the problem of detection of such a high-power laser, which might make entanglement generation with the $X$ measurement advantageous under the condition of the parameters in Table I.
Finesse can be enhanced in order to avoid this difficulty; however, it reduces the linear range of the optical cavity such that cavity length control becomes difficult.

\newpage


\begin{figure}[H]
    \centering
    \includegraphics[width=7.5cm]{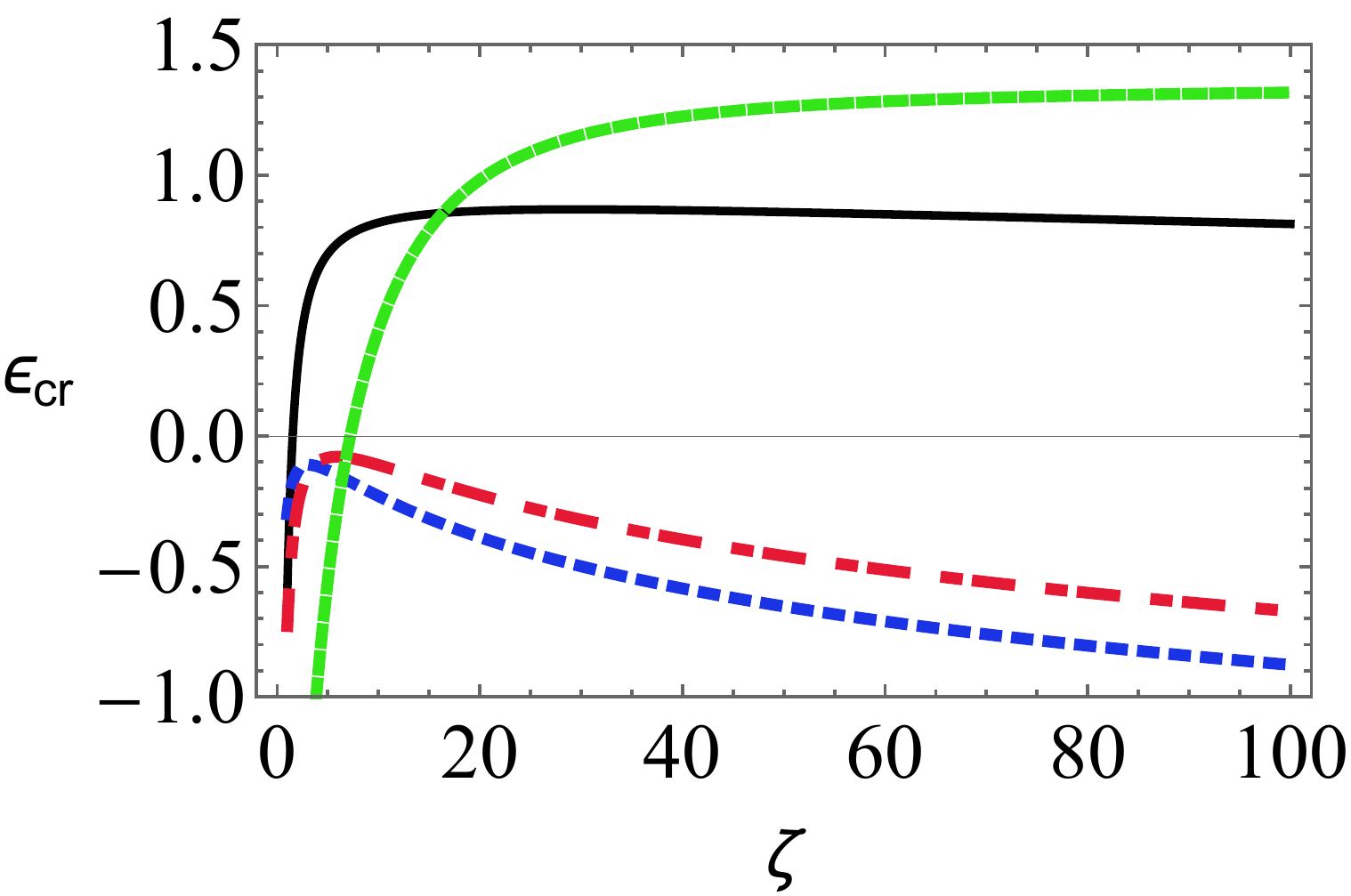}
    \hspace{1.5cm}
    \includegraphics[width=7.5cm]{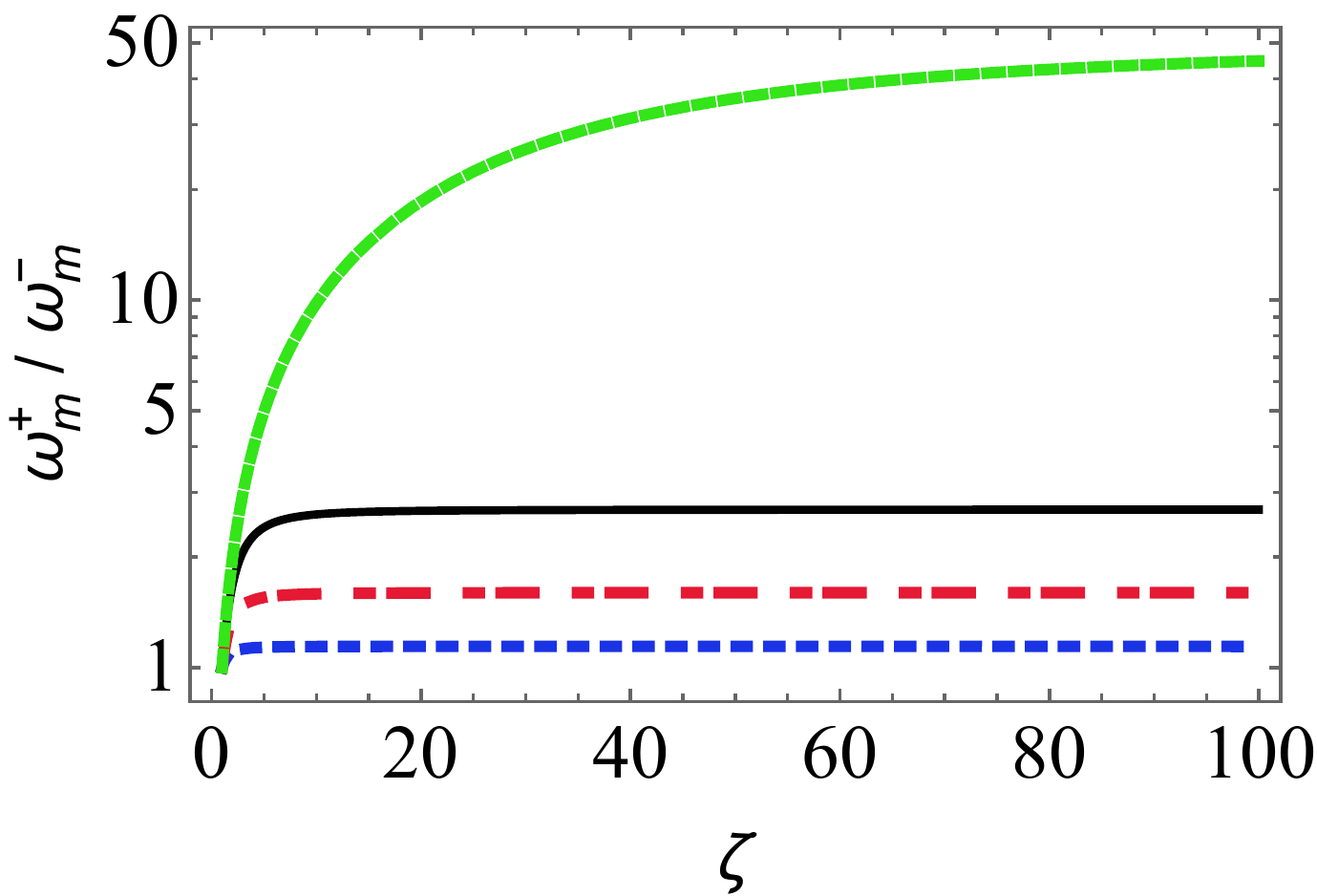}
    \caption{
    The behavior of the logarithmic negativity for the $X$ measurement (left panel) and the ratio of frequency of the mechanical common mode to that of the differential mode (right panel) as a function of $\zeta$. Each curve in the four colors assumes the same parameters as those of the circle in the same corresponding color in Fig.~\ref{fig:XEN}.
    }
    \label{fig:ZETA}
\end{figure}

\begin{figure}[H]
    \centering
    \includegraphics[width=7.5cm]{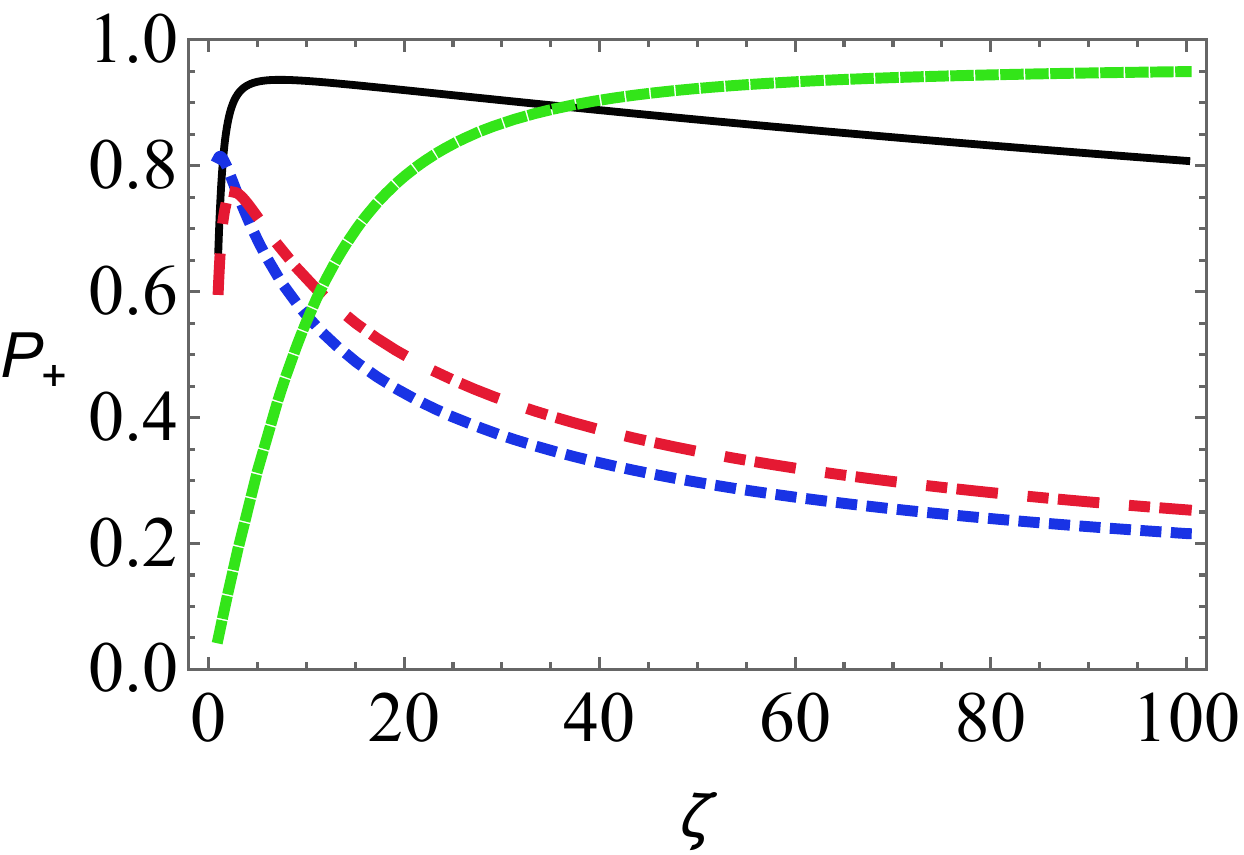}
    \caption{
    Purity of the mechanical common mode for the $X$ measurement as a function of $\zeta$. Here $P_-$ is fixed for each curved as $P_-=0.67$ (black solid curve) , $0.05$ (green solid curve), $0.81$ (blue dashed curve), and $0.60$ (red dash-dotted curve). 
    }
    \label{fig:pucomm}
\end{figure}

In the above analysis we fixed the parameter $\zeta=3$, which characterizes the asymmetry of the mechanical common and differential modes. Here we discuss how the entanglement behavior depends on the parameter $\zeta$. 
Figure \ref{fig:ZETA} shows the logarithmic negativity for the $X$ measurement (left panel) and the frequency ratio $\omega_{m}^{+}/\omega_{m}^{-}$ (right panel) 
as a function of $\zeta$.
The entanglement with the parameters in Table \ref{tab:parameter}, which is the solid black curve, saturates for $\zeta\simgt10$.
The green curve in the left panel of Fig.~\ref{fig:ZETA} increases significantly as $\zeta$ increases. 
We infer that the joint effect of the mechanical common mode and the differential mode on the entanglement is important. 
Fig.\ref{fig:pucomm} plots the purity of the common mode $P_+$ as functions of $\zeta$, where the purity of the differential model $P_-$ is fixed for each curve as $P_-=0.67$ (black solid curve) , $0.05$ (green solid curve), $0.81$ (blue dashed curve), and $0.60$ (red dash-dotted curve). 
The green curves in Fig. \ref{fig:pucomm} and the left panel of Fig. ~\ref{fig:ZETA} demonstrate that the entanglement appears by increasing the purity of the common mode when $\zeta$ increases, even when the purity of the differential mode is small. 
We note that this statement relies on the fact that the squeezing level of the differential mode is fixed.
Furthermore, the measurement efficiency, which is the coefficient of $q$ in the first term of the right hand side in Eq.~\eqref{Xi}, plays an important role for the squeezing through the detuning parameter when $\zeta$ changes.
Thus $\zeta$ is important for entanglement to control the purity and the asymmetry of the squeezing between the mechanical common and the differential modes which is caused by the asymmetric measurement efficiency.

\section{Summary and Conclusions}
We investigated the feasibility of generating a macroscopic Gaussian entanglement between mechanical oscillators coupled with cavity optical modes under continuous measurement and feedback control.
The mechanical oscillators are trapped with an optical spring owing to the detuning and squeezing achieved by measuring the output light.
We considered a Fabry--Perot--Michelson interferometer with a power-recycled mirror 
to generate asymmetry between the mechanical common and differential modes.
In this system, the two oscillators are entangled by the optical beams passing through the half-beam splitter.
This follows from the fact that
the entangled beam is generated by squeezed beams passing through a half-beam splitter.
In our optomechanical systems, the squeezed states of optical beams are produced through measurement with the Kalman filter, which optimizes the estimation of the oscillator quadratures, whose covariance matrix is determined by the Riccati equation in a steady state.
We derived the logarithmic negativity for the $X$ and $Y$ measurements in an analytic manner, including detuning and feedback control, and they are not limited to only the free-mass regions.

We analyzed the logarithmic negativity and phase space distribution, assuming 
tabletop experiments with the experimentally feasible parameters expected from the present technique \cite{Matsumoto20,MY}.
The quantum cooperativity $C_{\pm}/n_{\text{th}}^{\pm}$ and the detuning $\delta_{\pm}$ characterize the entanglement behavior.
The common mode and the differential mode of the oscillators are quantum-squeezed for $C_{\pm}/n_{\text{th}}^{\pm}\simgt1$, 
however, it is not enough for entanglement generation. Namely,
entanglement does not occur in the region with low purity even if both mechanical modes are squeezed.
Therefore, quantum squeezed states with high purity are necessary to generate entanglements.
The required values for generating the entanglement depend on the detuning and measurement schemes. 
For the $X$ measurement,
the condition of the quantum cooperativity 
$C_{-}/n_{\text{th}}^{-}\simgt3$
is required, assuming $\zeta=3$ and $\delta_-=0.2$.
These values will be achieved in the mid-term.
The required values for the $Y$ measurement can be slightly weakened depending on the level of detuning applied, though the homodyne ($Y$) measurement with high-power laser is experimentally difficult to achieve.
Thus, it is possible to experimentally generate quantum entanglement between mg-scale objects in the near future.
These predictions of quantum entanglement between macroscopic objects are not only a first step towards verifying the quantum nature of gravity but may also assist in verifying quantum mechanics in the macroscopic world.

For a realistic experimental setup, there are issues to be further considered. 
In the present analysis, coating thermal noise is ignored. 
This approximation is typically valid for the bandwidth around $1$ kHz
\cite{Matsumoto20}, but the influence of such a noise should be clarified in wide parameter regions. In an optomechanics with 
suspended mirrors, there exist additional mechanical modes other than the pendulum mode, e.g., rotation mode and violin modes \cite{Saulson}, which is also left for future investigations.

\acknowledgments

We are grateful for the discussions in the QUP theoretical collaboration. We especially thank S. Iso for his support and helpful discussions.
K.Y. was partially supported by JSPS KAKENHI, Grant No. 22H05263.
N.M. is supported by JSPS KAKENHI, Grant No. 19H00671 and JST FORESTO, Grant NO. JPMJFR202X.
D.M. is supported by JSPS KAKENHI, Grant No. 22J21267.
Y.S. was supported by the Kyushu University Innovator Fellowship in Quantum Science.
\appendix

\section{INPUT-OUTPUT RELATION FOR INTERFEROMETER}
We consider the input-output relation for the power-recycled Fabry-P\'{e}rot-Michelson interferometer shown in Fig.  \ref{fig:PRMI}.
The input-output relation of the individual Fabry-Perot cavity is obtained by
\begin{align}
    x_{\text{out}}^{1}&=
    x_{\text{in}}^{1}-\sqrt{\kappa}x_{1},\\
    x_{\text{out}}^{2}&=
    x_{\text{in}}^{2}-\sqrt{\kappa}x_{2},
\end{align}
where we assume that the optical decay rate is the same.
Since there is no mirror on the differential mode side, we obtain the output optical quadrature of the differential mode as
\begin{align}
    \label{ap:diff}
    x_{\text{out}}^{-}&=
    \frac{x_{\text{out}}^{1}-x_{\text{out}}^{2}}{\sqrt{2}}\notag\\
    &=\frac{x_{\text{in}}^{1}-x_{\text{in}}^{2}-\sqrt{\kappa}(x_{1}-x_{2})}{\sqrt{2}}\notag\\
    &=x_{\text{in}}^{-}-\sqrt{\kappa}x_{-}.
\end{align}
Then, we consider the input-output relationship considering the power-recycled mirror on the optical common mode side.
Using the transmissivity $T$ and the reflectivity $R=1-T$, we have
\begin{align}
    x_{\text{out}}^{+}
    &=\sqrt{R}x_{\text{in}}^{+}+\sqrt{T}\rho_{\text{out}}^{+},\notag\\
    \rho_{\text{in}}^{+}
    &=-\sqrt{R}\rho_{\text{out}}^{+}+\sqrt{T}x_{\text{in}}^{+},
\end{align}
where $\rho_{\text{out}}^{+}$ and $\rho_{\text{in}}^{+}$ are
\begin{align}
    \rho_{\text{out}}^{+}&=
    \frac{x_{\text{out}}^{1}+x_{\text{out}}^{2}}{\sqrt{2}},\\
    \rho_{\text{in}}^{+}&=
    \frac{x_{\text{in}}^{1}+x_{\text{in}}^{2}}{\sqrt{2}}.
\end{align}
Hence, the output optical quadrature of the common mode is
\begin{align}
    \label{ap:comm}
    x_{\text{out}}^{+}&=
    \frac{x_{\text{in}}^{1}+x_{\text{in}}^{2}-\sqrt{\frac{1-\sqrt{R}}{1+\sqrt{R}}\kappa}(x_{1}+x_{2})}{\sqrt{2}}\notag\\
    &=x_{\text{in}}^{+}-\sqrt{\frac{1-\sqrt{R}}{1+\sqrt{R}}\kappa}x_{+}.
\end{align}
Hence, we derive the relation between the optical decay rates of common mode and differential mode as
\begin{align}
    \kappa_{+}=\frac{1-\sqrt{R}}{1+\sqrt{R}}\kappa_{-}\equiv\frac{1}{\zeta}\kappa_{-}.
\end{align}

\section{LOGARITHMIC NEGATIVITY}
\label{LN}
Using the quality factor $Q_{\pm}=\omega_{m}^{\pm}/\gamma_{m}$ and cooperativity $C_{\pm}=4(g_{m}^{\pm})^{2}/\gamma_{m}\kappa_{\pm}$, we obtain
\begin{align}
    \frac{\lambda_{X}^{\pm}}{\gamma_{m}}&=
    \frac{16C_{\pm}\delta_{\pm}^{2}\eta}{(2\eta N_{\text{th}}+1)(1+4\delta_{\pm}^{2})^{2}}
    \equiv\lambda_{X}^{\pm\prime},\quad
    \frac{\lambda_{Y}^{\pm}}{\gamma_{m}}=
    \frac{4C_{\pm}\eta}{(2\eta N_{\text{th}}+1)(1+4\delta_{\pm}^{2})^{2}}
    \equiv\lambda_{Y}^{\pm\prime},\\
    \frac{\Lambda_{X}^{\pm}}{\gamma_{m}}&=
    -\frac{\Lambda_{Y}^{\pm}}{\gamma_{m}}=
    -\frac{8C_{\pm}\delta_{\pm}\eta}{(1+4\delta_{\pm}^{2})^{2}}\frac{2N_{\text{th}}+1}{2\eta N_{\text{th}}+1}
    \equiv\Lambda_{X}^{\pm\prime},\\
    \frac{\bar{n}_{\pm}}{\gamma_{m}}&=
    4n_{\text{th}}^{\pm}+2+\frac{4C_{\pm}}{(1+4\delta_{\pm}^{2})}(2N_{\text{th}}+1)
    \equiv\bar{n}_{\pm}^{\prime},\\
    \frac{\gamma_{I}^{\pm}}{\gamma_{m}}&=
    \sqrt{1-2Q_{\pm}^{2}
    \left(1+\frac{\Lambda_{I}^{\pm}}{\gamma_{m}Q_{\pm}}-\sqrt{1+2\frac{\Lambda_{I}^{\pm}}{\gamma_{m}Q_{\pm}}+\frac{\bar{n}_{\pm}\lambda_{I}^{\pm}}{\gamma_{m}^{2}Q_{\pm}^{2}}}\right)}\equiv\gamma_{I}^{\pm\prime},
\end{align}
where we define $\delta_{\pm}=\Delta/\kappa_{\pm}$.
From Eqs. \eqref{Sigma} and \eqref{detV}, we exactly derive the critical value as:
\begin{align}
    \epsilon_{\text{cr}}=
    -\frac{1}{2}\text{log}_{2}\Biggl[
    &\frac{(\gamma_{I}^{+\prime}-1)(\gamma_{I}^{-\prime}-1)}
    {4\lambda_{I}^{+\prime}\lambda_{I}^{-\prime}}
    \Biggl\{
     \left(
     \frac{(\gamma_{I}^{+\prime})^{2}+(\gamma_{I}^{-\prime})^{2}-\gamma_{I}^{+\prime}\gamma_{I}^{-\prime}-1}{Q_{+}Q_{-}}
     +2\frac{\Lambda_{I}^{+\prime}+Q_{+}}{Q_{-}}
     +2\frac{\Lambda_{I}^{-\prime}+Q_{-}}{Q_{+}}
     \right)\notag\\
    &-\Biggl(
     \frac{((\gamma_{I}^{+\prime})^{2}+(\gamma_{I}^{-\prime})^{2}-1)(\gamma_{I}^{+\prime}-\gamma_{I}^{-\prime})^{2}}{Q_{+}^{2}Q_{-}^{2}}
     +4\left(\frac{\Lambda_{I}^{+\prime}+Q_{+}}{Q_{-}}-\frac{\Lambda_{I}^{-\prime}+Q_{-}}{Q_{+}}\right)^{2}
     \notag\\
     &+4\frac{\gamma_{I}^{+\prime}(\gamma_{I}^{+\prime}-\gamma_{I}^{-\prime})(\Lambda_{I}^{+\prime}+Q_{+})}{Q_{+}Q_{-}^{2}}
     -4\frac{\gamma_{I}^{-\prime}(\gamma_{I}^{-\prime}-\gamma_{I}^{+\prime})(\Lambda_{I}^{-\prime}+Q_{-})}{Q_{+}^{2}Q_{-}}
     \Biggr)^{1/2}
    \Biggr\}
    \Biggr].
\end{align}

\section{SQUEEZING ANGLE}
The covariance matrix of a single mirror is diagonalized as
\begin{align}
   \bm V&=
    P^{-1}
    \left(\begin{array}{cc}
    \frac{1}{2}(V_{11}+V_{22}-\sqrt{(V_{11}-V_{22})^{2}+4V_{12}^{2}})&0\\
    0&\frac{1}{2}(V_{11}+V_{22}+\sqrt{(V_{11}-V_{22})^{2}+4V_{12}^{2}})
    \end{array}\right)
    P\notag\\
    &\equiv
    P^{-1}
    \left(\begin{array}{cc}
    E_{\text{min}}&0\\
    0&E_{\text{max}}
    \end{array}\right)
    P
    \end{align}
where $E_{\text{min}}~(E_{\text{max}})$ denotes the minimum (maximum) eigenvalue of the covariance matrix $\bm{\mathcal{V}}_{\pm}$.
$P$ is the rotation matrix
\begin{align}
    P&=
    \left(\begin{array}{cc}
    \sqrt{\frac{1}{2}\left(1+\frac{E_{\text{max}}+E_{\text{min}}-2V_{11}}{E_{\text{max}}-E_{\text{min}}}\right)}&\sqrt{\frac{1}{2}\left(1-\frac{E_{\text{max}}+E_{\text{min}}-2V_{11}}{E_{\text{max}}-E_{\text{min}}}\right)}\\
    -\sqrt{\frac{1}{2}\left(1-\frac{E_{\text{max}}+E_{\text{min}}-2V_{11}}{E_{\text{max}}-E_{\text{min}}}\right)}&\sqrt{\frac{1}{2}\left(1+\frac{E_{\text{max}}+E_{\text{min}}-2V_{11}}{E_{\text{max}}-E_{\text{min}}}\right)}
    \end{array}\right),
\end{align}
and its components are defined as
\begin{align}
    P&=
    \left(\begin{array}{cc}
    \cos(-\theta)&-\sin(-\theta)\\
    \sin(-\theta)&\cos(-\theta)
    \end{array}\right).
\end{align}
Hence, the squeezing angle can be obtained as follows:
\begin{align}
    \theta&=
    \arctan\left[\sqrt{\frac{\sqrt{(V_{22}-V_{11})^{2}+4V_{12}^{2}}+V_{11}-V_{22}}{\sqrt{(V_{22}-V_{11})^{2}+4V_{12}^{2}}-V_{11}+V_{22}}}\right]
    =\arctan\left[\sqrt{\frac{V_{11}-E_{\text{min}}}{E_{\text{max}}-V_{11}}}\right],
\end{align}

\end{document}